\documentclass[twocolumn,showpacs,superscriptaddress,preprintnumbers,amsmath,amssymb,longbibliography,pre]{revtex4-1}
\usepackage{graphicx}
\usepackage{dcolumn}
\usepackage{bm}
\usepackage{amsmath}
\usepackage{color}

\begin{document}

\title{Critical scaling near the yielding transition in granular media}
\author{Abram H. Clark}
\affiliation{Department of Physics, Naval Postgraduate School, Monterey, California 93943, USA}
\affiliation{Department of Mechanical Engineering and Materials Science, Yale University, New Haven, Connecticut 06520, USA}
\author{Jacob D. Thompson}
\affiliation{Department of Physics, Naval Postgraduate School, Monterey, California 93943, USA}
\author{Mark D. Shattuck}
\affiliation{Benjamin Levich Institute and Physics Department, The City College of the City University of New York, New York, New York 10031, USA}
\author{Nicholas T. Ouellette}
\affiliation{Department of Civil and Environmental Engineering, Stanford University, Stanford, California 94305, USA}
\author{Corey S. O'Hern}
\affiliation{Department of Mechanical Engineering and Materials Science, Yale University, New Haven, Connecticut 06520, USA}
\affiliation{Department of Physics, Yale University, New Haven, Connecticut 06520, USA}
\affiliation{Department of Applied Physics, Yale University, New Haven, Connecticut 06520, USA}

\begin{abstract}
We show that the yielding transition in granular media displays second-order critical-point scaling behavior. We carry out discrete element simulations in the low inertial number limit for frictionless, purely repulsive spherical grains undergoing simple shear at fixed nondimensional shear stress $\Sigma$ in two and three spatial dimensions. To find a mechanically stable (MS) packing that can support the applied $\Sigma$, isotropically prepared states with size $L$ must undergo a total strain $\gamma_{\rm ms}(\Sigma,L)$. The number density of MS packings ($\propto \gamma_{\rm ms}^{-1}$) vanishes for $\Sigma > \Sigma_c \approx 0.11$ according to a critical scaling form with a length scale $\xi \propto |\Sigma - \Sigma_c|^{-\nu}$, where $\nu \approx 1.7-1.8$. Above the yield stress ($\Sigma>\Sigma_c$), no MS packings that can support $\Sigma$ exist in the large system limit, $L/\xi \gg 1$. MS packings generated via shear possess anisotropic force and contact networks, suggesting that $\Sigma_c$ is associated with an upper limit in the degree to which these networks can be deformed away from those for isotropic packings.
\end{abstract}

\date{\today}


\maketitle	

\section{Introduction}
Granular materials consist of macroscopic grains that interact via dissipative contact forces. Their response to external forcing depends on the ratio $\Sigma = \tau/p$ of the applied shear stress $\tau$ to the normal stress $p$, where $p$ is small compared to the stiffness of the grains~\cite{dacruz2005,jop2006}. Granular media, like other amorphous materials~\cite{gardiner1998,yoshimura1987,coussot2002,boromand2017}, possess a yield stress $\Sigma_c$. Generally, grains will always rearrange when the applied forces are changed. However, when $\Sigma<\Sigma_c$, grains move temporarily until finding a solid-like mechanically stable (MS) packing that can support the applied $\Sigma$~\cite{toiya2004,xu2006,peyneau08}. When $\Sigma = \Sigma_c$, the strain $\gamma_{\rm ms}$ required to find MS packings diverges. When $\Sigma>\Sigma_c$, grains cannot find MS packings, and fluid-like flow persists indefinitely.

In the jamming paradigm~\cite{ohern2003,donev2004,olsson2007,vanhecke2009,Tighe2010,Nordstrom2010,olsson2011}, which is commonly used to understand fluid-solid transitions in granular materials, the packing fraction $\phi$ is the controlling variable. Fluid- and solid-like states occur for $\phi<\phi_J$ and $\phi>\phi_J$, respectively. A diverging length scale $\xi_J \propto |\phi - \phi_J|^{-\nu_J}$ controls the mechanical response near $\phi_J$~\cite{olsson2007,Tighe2010,Nordstrom2010,olsson2011}. However, MS packings of frictionless grains at fixed $p$ and varied $\Sigma$ all have a packing fraction $\phi_{\rm ms}(\Sigma) \approx \phi_J$~\cite{peyneau08}. Thus, $\Sigma \approx \Sigma_c$ may represent a fluid-solid transition distinct from jamming, where the structure of the force and contact networks, not $\phi$, plays a dominant role.

In this paper, we show evidence that the number density of MS packings vanishes at $\Sigma= \Sigma_c$ in the large-system limit, with second-order critical scaling that is not related to $\phi$ but instead to the structure of the force and contact networks. We measure $\gamma_{\rm ms}$ in systems of frictionless grains subjected to simple shear as a function of $\Sigma$ and system size $L$. We postulate a second-order critical point scaling form for $\gamma_{\rm ms}$ with a diverging length scale $\xi \propto |\Sigma - \Sigma_c|^{-\nu}$. The data for $\gamma_{\rm ms}$ collapse onto two branches: $\Sigma>\Sigma_c$ and $\Sigma<\Sigma_c$. For simple shear in two (2D) and three dimensions (3D), we find $\Sigma_c\approx 0.11$, in agreement with previous studies~\cite{peyneau08,dacruz2005,kamrin2014}. MS packings exist for $\Sigma>\Sigma_c$ in small systems, but the number vanishes as $L/\xi$ increases. For $\Sigma < \Sigma_c$, MS packings exist for all $L$, and large systems ($L>\xi$) are equivalent to compositions of uncorrelated smaller systems. Our results are insensitive to changes in the boundary conditions and driving method, which we explicitly show by performing additional simulations in a riverbed-like geometry in the viscous or slow-flow limit~\cite{clark2015hydro,clark2017}. 

We find that the packing fraction $\phi_{\rm ms}(\Sigma,L)$ of MS packings is nearly independent of $\Sigma$. However, the anisotropy in both the stress and contact fabric tensors of MS packings increases with $\Sigma$, suggesting that $\Sigma_c$ is associated with an upper limit to the structural anisotropy that can be realized in the large-system limit~\cite{peyneau08}. These results may help explain recent studies~\cite{kamrin2012,bouzid2013,henann2014,kamrin2015,bouzid2015} showing that accurately modeling granular flows requires a cooperative length scale that grows as a power law in $\Sigma-\Sigma_c$. Our results may also be relevant to other amorphous solids that show similar spatial cooperativity near yielding~\cite{coussot2002,bocquet2009,karmakar2010,lin2014}.

The remainder of the manuscript is organized as follows. In Sec.~\ref{sec:methods}, we describe our simulation methods. In Sec.~\ref{sec:results} we present our results, including the critical scaling of $\gamma_{\rm ms}$ in Sec.~\ref{sec:scaling} and the microstructural properties of MS packings in Sec.~\ref{sec:microstructure}. Section~\ref{sec:conclusion} contains a summary and conclusions. We include additional details in Appendix~\ref{Appendix-EOM} on the equations of motion and dimensional analysis. Appendix~\ref{Appendix-crit-vals} demonstrates our methods for determining the critical exponents and $\Sigma_c$. Appendix~\ref{Appendix-L-xi} gives further discussion on the scaling collapse of $\langle \gamma_{\rm ms} \rangle^{-1} $ versus $\Sigma$.

\begin{figure*}
\raggedright \hspace{5mm}  (a) \hspace{50mm} (b) \hspace{45mm} (c)
\\ \centering
\includegraphics[trim=20mm 15mm 12mm 20mm, clip, width=0.34\textwidth]{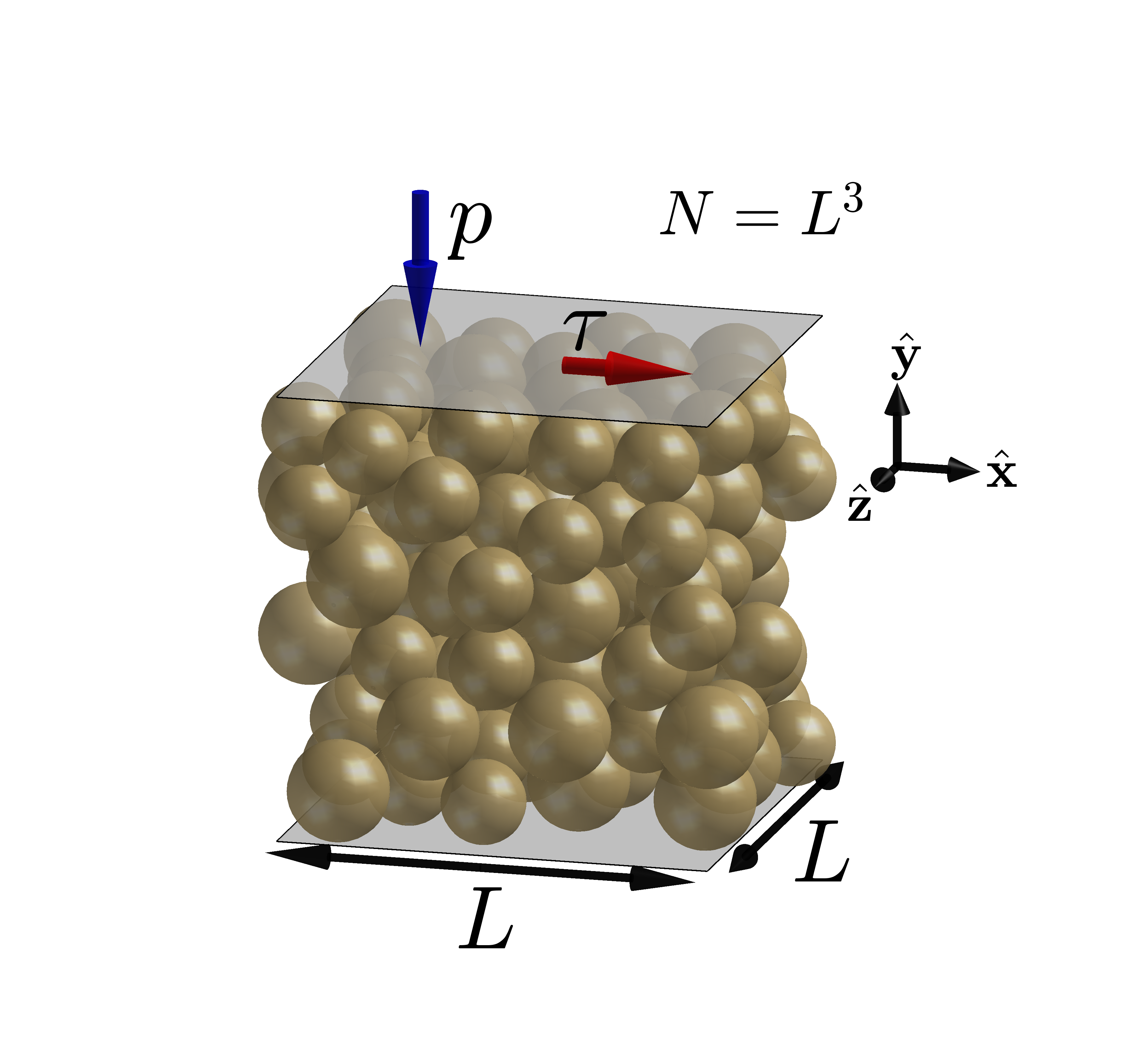}
\includegraphics[trim=10mm 0mm 5mm 0mm, clip, width=0.26\textwidth]{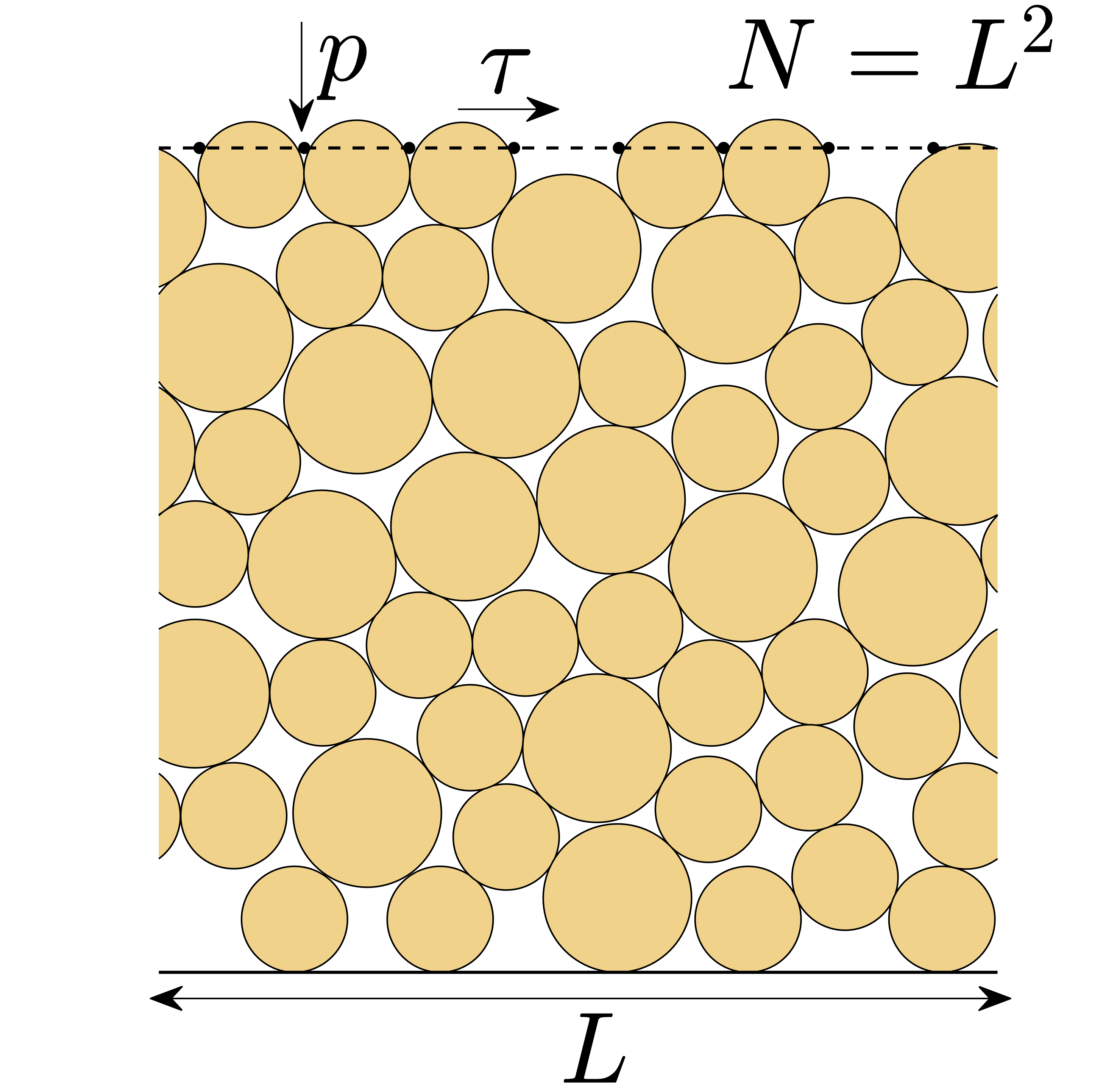} 
\includegraphics[trim=0mm 0mm 10mm 5mm, clip, width=0.36\textwidth]{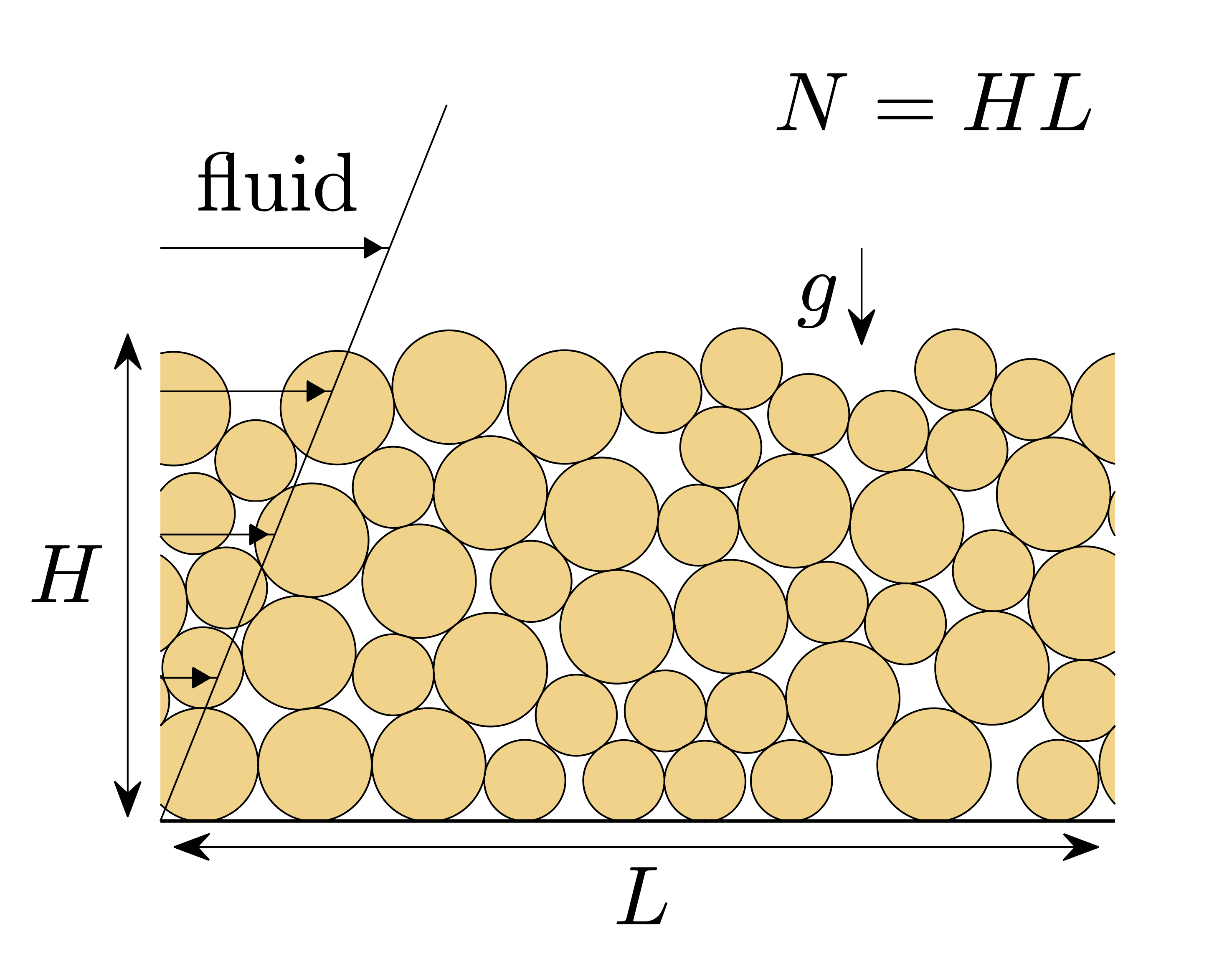} 
\raggedright
\caption{Schematics of the simulation procedure for (a) 3D simple shear and (b) 2D simple shear. In (a) and (b), MS packings are first created under only a fixed normal force per area $-p\hat{y}$. We then apply a shear force per area $\tau\hat{x}$ and search for an MS packing at a given $\Sigma=\tau/p$ and length $L$ in units of the small grain diameter $D$. (c) A schematic of the simulation procedure for the riverbed-like geometry. MS packings are first created via sedimentation under gravity $-g \hat{y}$, then driven by a fluid-like drag force in the $x-$direction.}
\label{fig:cartoon}
\end{figure*}


\section{Methods} 
\label{sec:methods}

As depicted in Fig.~\ref{fig:cartoon}, we perform discrete element method simulations of simple shear in 3D and 2D, as well as in a 2D riverbed-like geometry subjected to a linear flow profile in the viscous limit. For the simple shear simulations, we study systems of bidisperse frictionless spheres in 3D and disks in 2D. Two-thirds of the grains are small and one-third are large, with diameter ratio 1.2 in 3D and 1.4 in 2D. The lateral directions $x$ (in 2D and 3D) and $z$ (in 3D only) are periodic with length $L$, where $L$ is the length of the box edge in units of the small grain diameter $D$. The fixed lower $y$ boundary consists of a no-slip wall. The system is driven by the upper boundary, which is a plate consisting of rigidly connected small particles, with gaps that are large enough to prevent slip and small enough to stop bulk grains from passing through the plate. We have checked that our results are insensitive to the details of the top plate, provided no slip occurs between the plate and grains. We apply downward force per area $-p\hat{y}$ and horizontal force per area $\tau \hat{x}$ to the upper plate and solve Newton's equations of motion for the wall as well as $N$ grains  using a modified velocity Verlet integration scheme. In 3D, we vary $N=L^3$ from $L=3$, $N=27$ to $L=16$, $N=4096$. In 2D, we vary $N=L^2$ from $L=7$, $N=49$ to $L=40$, $N=1600$. Grains interact via purely repulsive, linear springs with force constant $K$. For the systems driven by simple shear, we include a viscous damping force $-B\mathbf{v}$ in the equations of motion for the top plate and $N$ grains, where $\mathbf{v}$ is the absolute velocity and $B$ is the damping coefficient.

The equations of motion for simple shear, described in detail in Appendix \ref{Appendix-EOM}, are governed by three nondimensional parameters:
\begin{align}
\Gamma &= \frac{B}{\sqrt{mpD^{d-2}}}, \\
\kappa &= \frac{K}{pD^{d-1}}, \\
\Sigma &= \frac{\tau}{p},
\label{eq:gamma-kappa-sigma}
\end{align}
where $d$ is the spatial dimension. $\Gamma$ is the dimensionless damping parameter, which we set equal to 5, and $\kappa$ is a dimensionless grain stiffness. We set $\kappa = 10^3$, meaning that $\phi \approx \phi_J(L) + 0.001$, where $\phi_J(L)$ is the jamming packing fraction at a given $L$. Our results are insensitive to $\kappa$ in this limit, which we verify for several values of $\kappa > 200$. We set $\Gamma = 5$, which maintains an inertial number $I=\dot{\gamma}\sqrt{m/p}<10^{-4}$ (where $\dot{\gamma}$ is the strain rate) in the slow- or creep-flow limit, $I<10^{-3}$\cite{dacruz2005,kamrin2012}. We control force and not $\dot{\gamma}$, so there are fluctuations in $I$, but $\Gamma = 5$ keeps $I<10^{-4}$ even for $\Sigma>\Sigma_c$. We have explicitly checked that our results are independent of $\Gamma$ for several values of $\Gamma \geq 3$. 

For the simple shear simulations, initial states ($\Sigma=0$) are prepared via uniaxial compression. Specifically, we begin with the top plate at very large $y\gg L$, and we place the grains sparsely throughout the domain between the top plate and the lower boundary. We then apply finite $p$ to the top plate and allow it to move freely until an MS packing is found. We then apply finite $\Sigma$ to the top plate, which can move in all directions. The simulation ends when the upward and horizontal forces from the grains acting on the top plate exactly balance the applied $\Sigma$. We find similar results when $\Sigma$ is increased incrementally in small steps, and the total strain is integrated. Average grain displacement profiles are linear for both $\Sigma<\Sigma_c$ and $\Sigma>\Sigma_c$~\cite{xu2005,xu2006}, as expected for this system. 

In addition to simple shear, we study 2D systems of bidisperse frictionless grains in a riverbed-like geometry, depicted in Fig.~\ref{fig:cartoon}(c), which is similar to the system studied in Ref.~\cite{clark2015hydro}. The domain has a no-slip lower boundary at $y=0$, a free upper boundary, and periodic horizontal boundaries in the $x-$direction with length $L$ (in units of the small grain diameter $D$). We use $N=5L$ such that the system has height $H\approx 5D$. We vary $L$ and $N$ between $L=3$, $N=15$ and $L=320$, $N=1600$. Grains interact via purely repulsive, linear springs with force constant $K$. We apply a buoyancy-reduced gravitational force $-m g'\mathbf{\hat{y}}$ and a horizontal fluid force $B(v_0 y_i/H \mathbf{\hat{x}} - \mathbf{v}_i)$ to each grain $i$, where $B$ is a drag coefficient, $y_i$ is the height above the lower boundary, $v_0$ is the characteristic velocity at the bed surface, and $\mathbf{v}_i$ is the grain velocity. We find similar results for several different fluid flow profiles. The equations of motion, shown in Appendix \ref{Appendix-EOM}, are again governed by three dimensionless parameters
\begin{align}
\Gamma' &= \frac{B/m}{\sqrt{g'/D}}, \\
\kappa' &= \frac{K}{mg'}, \\
\Sigma' &= \frac{Bv_0}{mg'}.
\label{eqn:nondimen-params}
\end{align}
We again set $\Gamma' = \frac{B/m}{\sqrt{g'/D}}=5$ and $\kappa' = \frac{K}{mg'} = 1000$ and vary the dimensionless shear stress $\Sigma' = \frac{Bv_0}{mg'}$. Our results are again independent of $\kappa'$ and $\Gamma'$ in this regime. We prepare beds via sedimentation with $\Sigma' = 0$ and then apply finite $\Sigma'$ and allow the system to evolve until the system stops at an MS packing.

\section{Results}
\label{sec:results}

\subsection{Critical scaling of $\gamma_{\rm ms}$}
\label{sec:scaling}

\begin{figure}
\raggedright
(a) \hspace{39mm} (b) \\ \centering
\includegraphics[trim=0mm 0mm 5mm 0mm, clip, width=0.49\columnwidth]{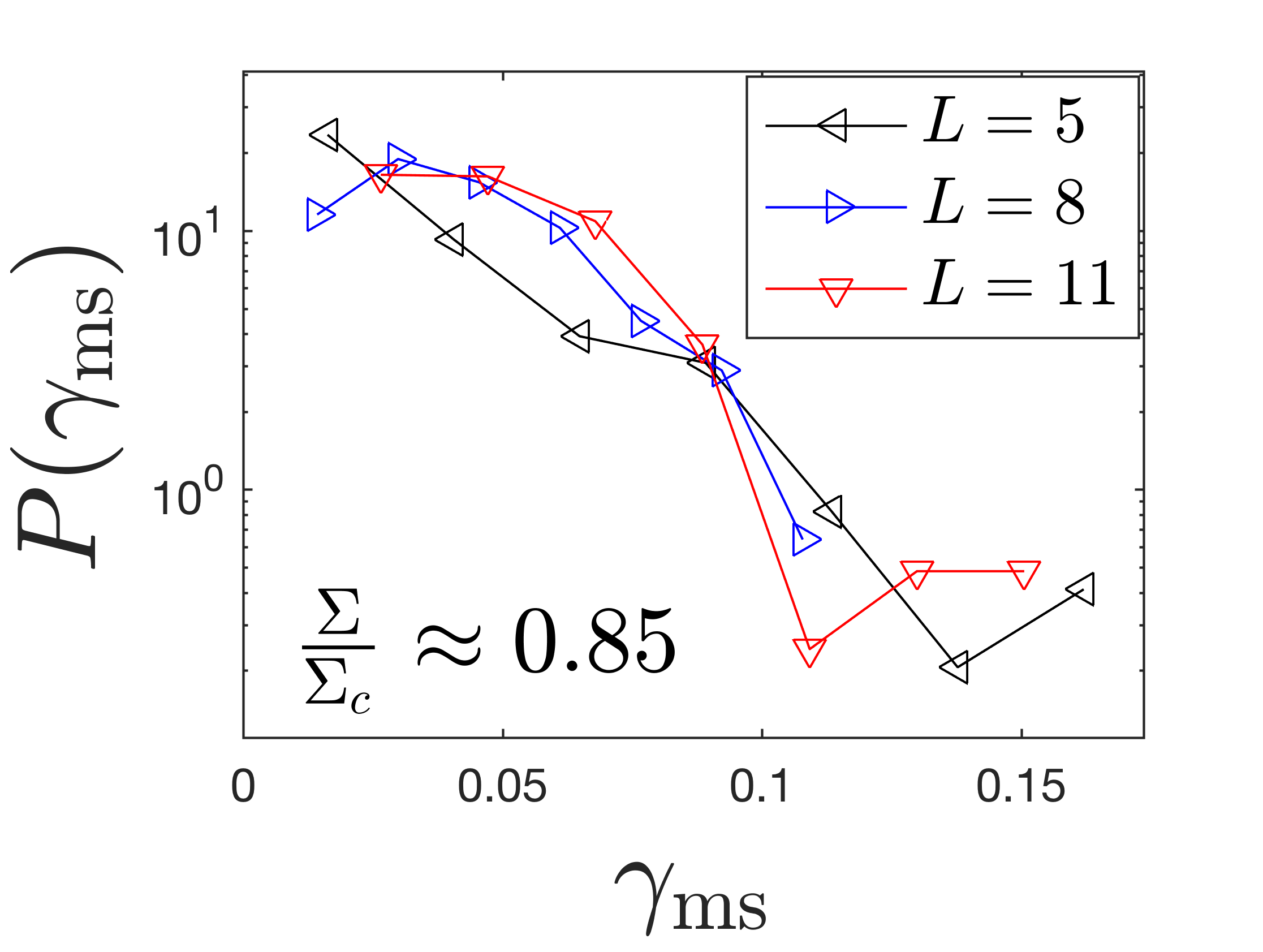}
\includegraphics[trim=0mm 0mm 5mm 0mm, clip, width=0.49\columnwidth]{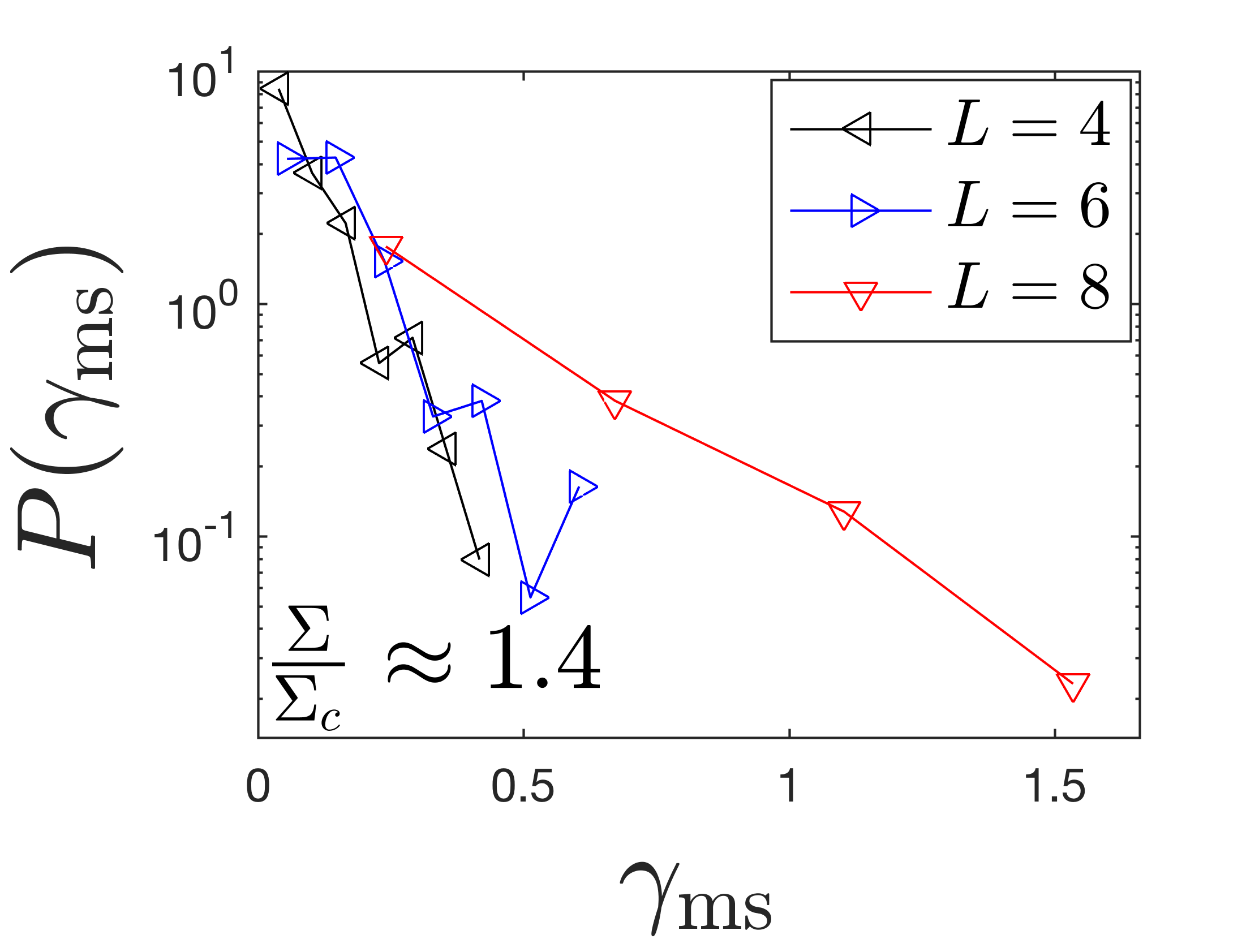} \\
\raggedright (c) \\ \centering
\includegraphics[trim=0mm 0mm 10mm 5mm, clip, width=0.9\columnwidth]{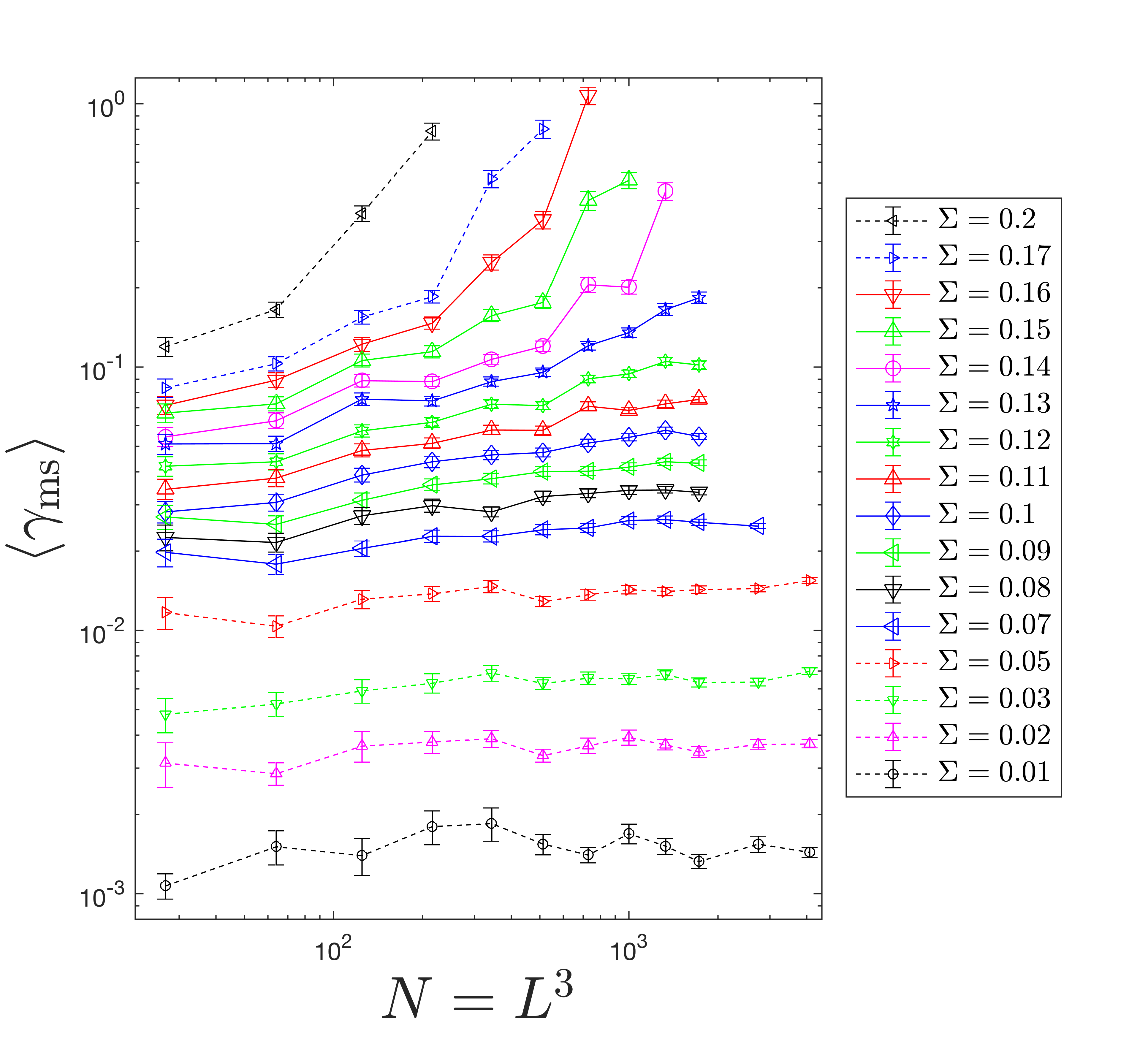} \\
\raggedright (d) \\ \centering
 \includegraphics[trim=0mm 0mm 10mm 5mm, clip, width=0.9\columnwidth]{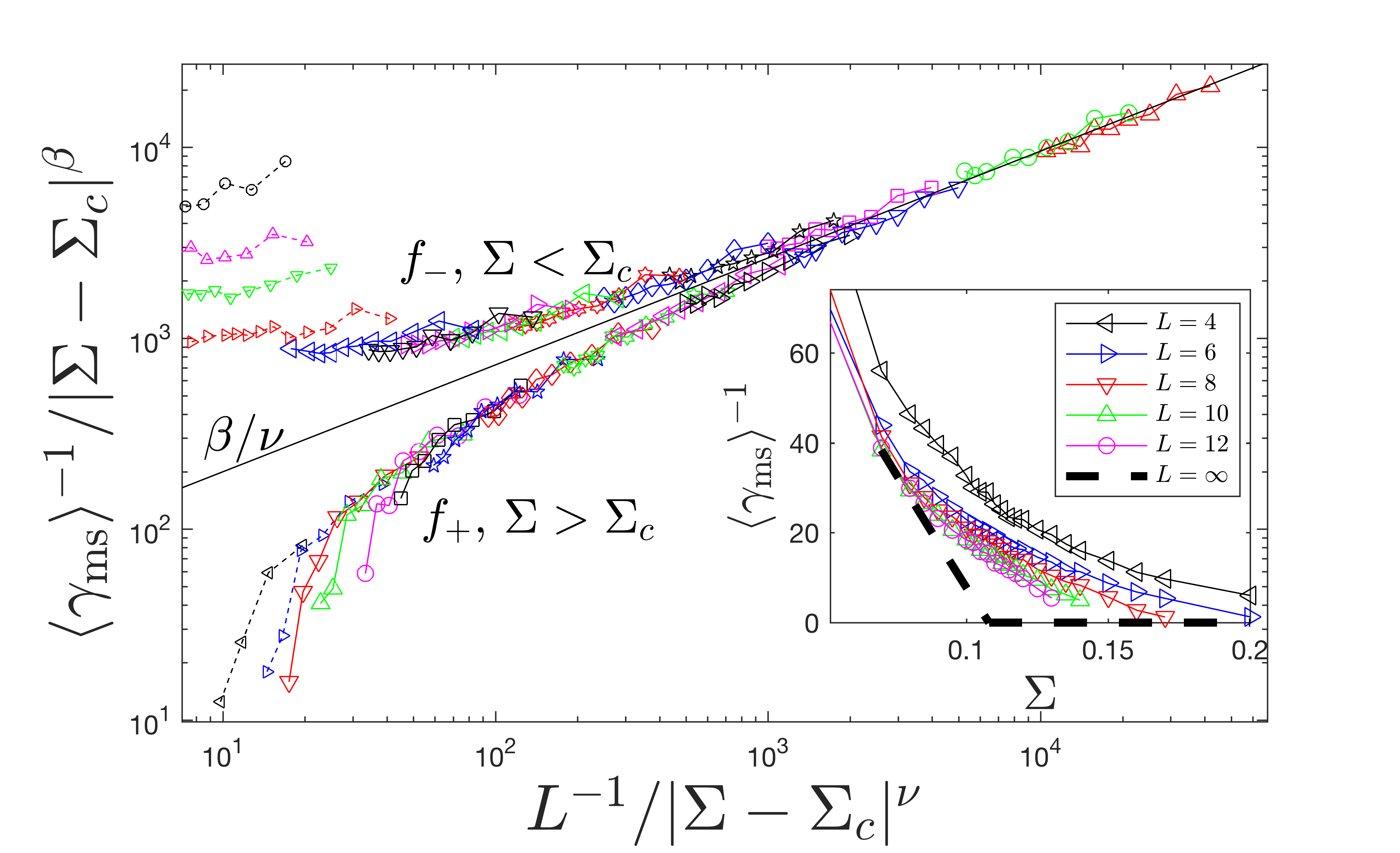} 
\caption{(a-b) Distributions
$P(\gamma_{\rm ms})$ of the strain $\gamma_{\rm ms}$ in 3D simple shear between the initial and final MS packings for (a) $\Sigma<\Sigma_c$, (b) $\Sigma>\Sigma_c$, and several $L$. (c) The mean strain $\langle \gamma_{\rm ms} \rangle$ in 3D simple shear between initial and final MS packings plotted versus system size $N=L^3$ for several values of applied stress $\Sigma$. Solid (dashed) lines correspond to $|\Sigma-\Sigma_c|/\Sigma_c$ less (greater) than 0.5. Error bars are the standard error of the mean, given by the standard deviation within the sample divided by the square root of the number of trials. (d) The data from panel (c), plus additional data for more $\Sigma$, collapses when using the scaled variables $\langle \gamma_{\rm ms}\rangle^{-1}/|\Sigma - \Sigma_c|^\beta$ and $L^{-1}/|\Sigma - \Sigma_c|^\nu$, where $\beta/\nu=0.56$, $\nu=1.7$, and $\Sigma_c = 0.109$. The inset shows $\langle \gamma_{\rm ms}(\Sigma,L)\rangle^{-1}$ versus $\Sigma$ for different $L$. The dashed line gives the large-system limit implied by Eq.~\eqref{eq:scaling-1} and the main plot in (d). The data in the inset is at constant $L$, not constant $L^{-1}/|\Sigma - \Sigma_c|^\nu$; see text and Appendix~\ref{Appendix-L-xi} for further discussion.}
\label{fig:gamma-ms-3D}
\end{figure}

We begin with results for simple shear in 3D. We define the shear strain $\gamma_{\rm ms}$ as the total distance the top plate moves in the $x$-direction divided by the average of the initial and final $y$-positions of the top plate. In Fig.~\ref{fig:gamma-ms-3D}(a) and (b), we show the distribution $P(\gamma_{\rm ms})$ for two illustrative values of $\Sigma$ over a range of system sizes $L$, obtained using $200$ simulations for each $L$. For small $L$ above and below $\Sigma_c$, the distributions are roughly exponential, $P(\gamma_{\rm ms}) \approx \langle \gamma_{\rm ms} \rangle^{-1} \exp(-\gamma_{\rm ms}/\langle{\gamma}_{\rm ms}\rangle)$. This form indicates an underlying physical process resembling absorption~\cite{bertrand2016}, where objects propagate through space and each stops whenever it encounters an absorber. For absorption processes, the propagation distance distributions are exponential, as in Fig.~\ref{fig:gamma-ms-3D}, and the mean ``travel distance'' is inversely proportional to the density of absorbers. For sheared packings, the mean travel distance is $\langle{\gamma}_{\rm ms}(\Sigma,L)\rangle$. Thus, we use $\langle{\gamma}_{\rm ms}(\Sigma,L)\rangle^{-1}$ as a measure of the number density of MS packings.

In Fig.~\ref{fig:gamma-ms-3D} (c), we plot $\langle \gamma_{\rm ms} \rangle$ versus $L$ over a range of $\Sigma$. Figure~\ref{fig:gamma-ms-3D} (d) shows that these data can be collapsed by plotting the scaled variables $L^{-1}/|\Sigma - \Sigma_c|^\nu$ and $\langle \gamma_{\rm ms}\rangle^{-1}/|\Sigma - \Sigma_c|^\beta$. This collapse implies that finite size effects for $\langle \gamma_{\rm ms}(\Sigma,L)\rangle^{-1}$ depend on a diverging correlation length $\xi \propto |\Sigma - \Sigma_c|^{-\nu}$, 
\begin{equation}
\langle \gamma_{\rm ms}(\Sigma,L) \rangle^{-1} = |\Sigma - \Sigma_c|^\beta f_\pm\left(\frac{L^{-1}}{|\Sigma - \Sigma_c|^\nu}\right).
\label{eq:scaling-1}
\end{equation}
Here, $f_\pm$ are the critical scaling functions for $\Sigma > \Sigma_c$ and $\Sigma < \Sigma_c$, respectively, which capture the finite-size effects. Note that all quantities in Eq.~\eqref{eq:scaling-1} are dimensionless. As shown in Appendix~\ref{Appendix-crit-vals}, we determine the critical values by fitting the data to this functional form, where the critical values are fit parameters. We systematically exclude small system sizes and large deviations $|\Sigma - \Sigma_c|$. We quantify the quality of the fits using the reduced chi-squared metric, $\chi^2 = \sum_i (\Delta_i)^2/e_i^2$, where the sum is over all data points $i$ used in the fit, $\Delta_i$ is the difference between the data and the fit, and $e_i$ is the standard error in the mean (i.e., the standard deviation within that sample divided by the square root of the number of trials), represented as error bars in Fig.~\ref{fig:gamma-ms-3D}(c). We search for fits where $\chi^2/n \approx 1$~\cite{olsson2011}, where $n$ is the number of data points minus the number of fit parameters, and the critical values are independent of the range of $|\Sigma - \Sigma_c|$. From this analysis, shown in Appendix~\ref{Appendix-crit-vals}, we estimate  $\nu = 1.7 \pm 0.5$, $\Sigma_c = 0.109 \pm 0.005$ and $\beta / \nu = 0.57 \pm 0.07$. The uncertainty ranges represent the scatter in the fit results plus one standard deviation. Despite the uncertainty, $\nu \approx 1.7$ for yielding appears distinct from $\nu_J\approx 0.6-1$ for jamming~\cite{ohern2003,olsson2007,olsson2011}, suggesting that these are two separate, though possibly related, zero-temperature transitions. 

The inset in Fig.~\ref{fig:gamma-ms-3D}(d) shows $\langle \gamma_{\rm ms} \rangle^{-1}$ plotted versus $\Sigma$ for different $L$, as well as the large-system limit (dashed, black line) implied by the scaling in the main panel of Fig.~\ref{fig:gamma-ms-3D}(d). For $\Sigma<\Sigma_c$, $f_-$ becomes constant at small $L^{-1}/|\Sigma - \Sigma_c|^\nu$ (\textit{i.e.}, $L>\xi$). This means that, in the large-system limit, $\langle \gamma_{\rm ms} \rangle^{-1}$ vanishes nonanalytically at $\Sigma = \Sigma_c$ according to $\langle \gamma_{\rm ms} \rangle^{-1} \propto |\Sigma-\Sigma_c|^{\beta}$. Also at small $L^{-1}/|\Sigma - \Sigma_c|^\nu$ for $\Sigma < \Sigma_c$, a peak develops in $P(\gamma_{\rm ms})$ at $\gamma_{\rm ms} > 0$, as shown in Fig.~\ref{fig:gamma-ms-3D}(a). We interpret this behavior as spatial decorrelation, where large systems behave like compositions of uncorrelated exponentially distributed random variables, yielding a distribution that is peaked at $\gamma_{\rm ms}>0$ with a mean that is independent of $L/\xi$. For $\Sigma > \Sigma_c$, $f_+$ is finite but tends to zero for small $L^{-1}/|\Sigma - \Sigma_c|^\nu$. This means that the number of MS packings vanishes for $\Sigma>\Sigma_c$ as $L/\xi$ increases. If $f_+$ approaches a vertical asymptote, MS packings do not exist for $\Sigma>\Sigma_c$ and finite $L/\xi$. Otherwise, MS packings only vanish for infinite $L/\xi$. Further studies with larger system sizes are required to address this specific point.

Note that the data we present in the inset to Fig.~\ref{fig:gamma-ms-3D}(d) approach the $L/\xi \rightarrow \infty$ limiting form (dashed curve) only for $\Sigma < \Sigma_c$ and for $\Sigma > \Sigma_c$, but not near $\Sigma_c$. The data does not collapse near $\Sigma_c$ because the scaled system size $L^{-1}/|\Sigma-\Sigma_c|^\nu$ changes significantly as $\Sigma$ is varied at fixed $L$. We show the data in the inset to Fig.~\ref{fig:gamma-ms-3D}(d) at constant $L^{-1}/|\Sigma-\Sigma_c|^\nu$ in Appendix~\ref{Appendix-L-xi}.   


Figure~\ref{fig:gamma-ms-2D} shows that the results for 2D systems with boundary driven simple shear are similar to those in 3D. Distributions for $P(\gamma_{\rm ms})$ (not shown) are similar to the 3D case, which are shown in Fig.~\ref{fig:gamma-ms-3D}(a) and (b). In Fig.~\ref{fig:gamma-ms-2D}(a), we plot $\langle \gamma_{\rm ms} \rangle$ versus $N=L^2$ for selected values of $\Sigma$. Figure~\ref{fig:gamma-ms-2D}(b) shows that these data (plus additional data) collapse by plotting the scaled variables $L^{-1}/|\Sigma - \Sigma_c|^\nu$ and $\langle \gamma_{\rm ms}\rangle^{-1}/|\Sigma - \Sigma_c|^\beta$. Using a similar fitting analysis to that described above for 3D systems undergoing simple shear, we obtain $\nu = 1.84 \pm 0.3$, $\Sigma_c = 0.11 \pm 0.01$, and $\beta/\nu \approx 0.57 \pm 0.06$. 

\begin{figure}
\raggedright
(a) \\  \includegraphics[trim=0mm 0mm 0mm 0mm, clip, width=\columnwidth]{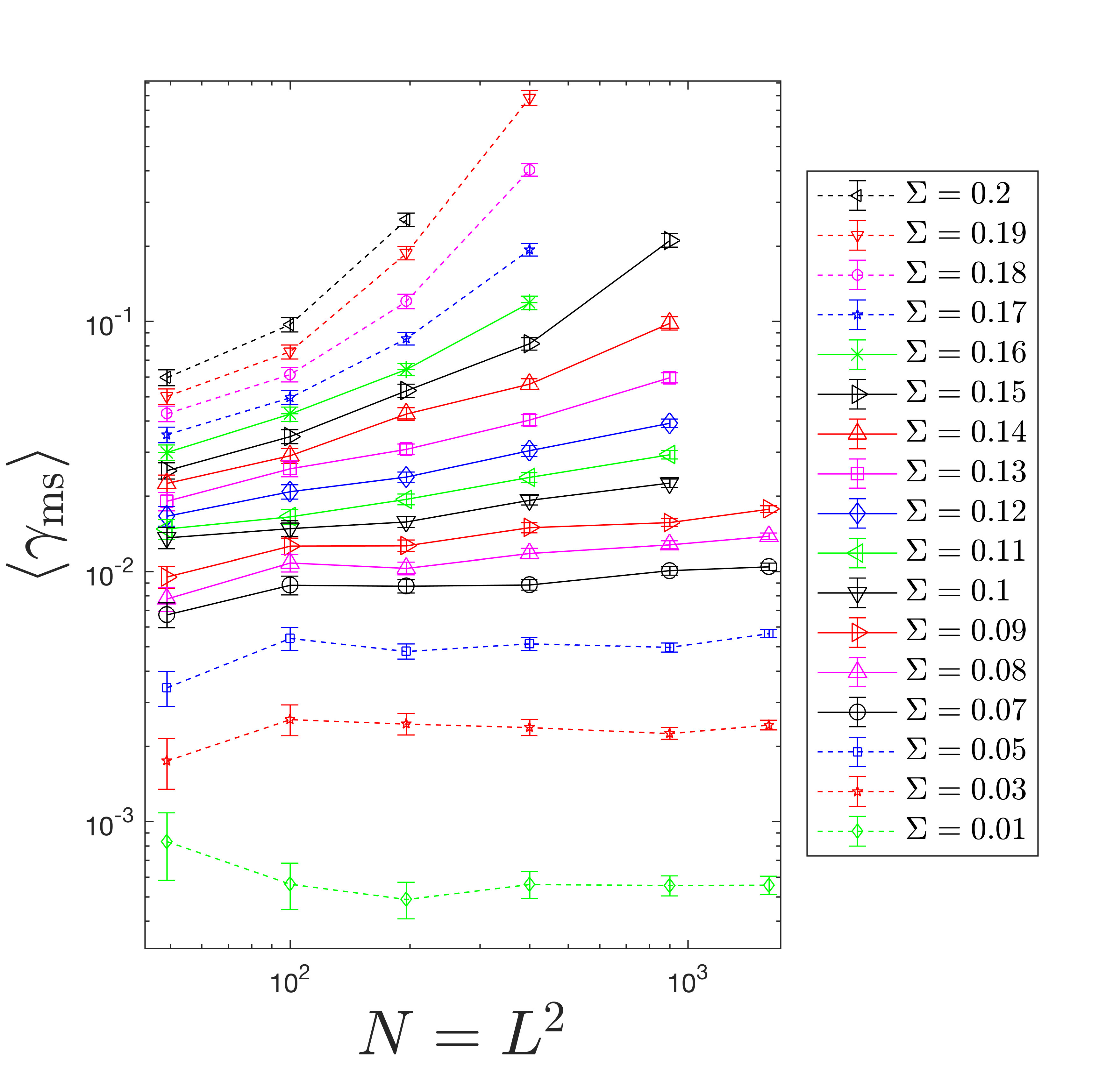} \\
(b) \\  \includegraphics[trim=0mm 0mm 0mm 0mm, clip, width=\columnwidth]{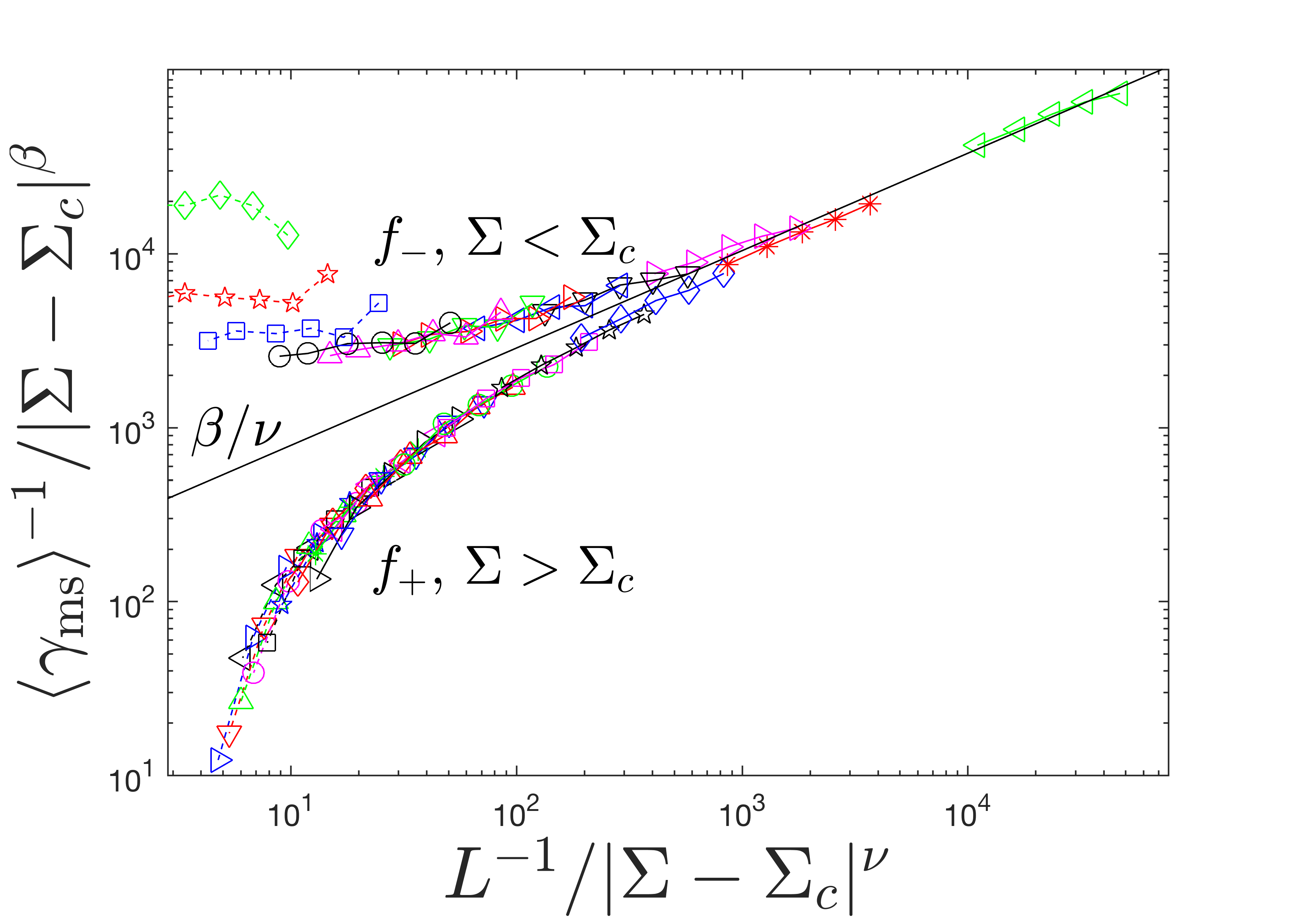} \\
\caption{(a) The mean strain $\langle \gamma_{\rm ms} \rangle$ in 2D simple shear between initial and final MS packings plotted versus system size $N=L^2$ for several values of applied stress $\Sigma$. Solid (dashed) lines correspond to $|\Sigma-\Sigma_c|/\Sigma_c$ less (greater) than 0.5. Error bars are the standard error of the mean, given by the standard deviation within the sample divided by the square root of the number of trials. (b) The data from panel (a), plus additional data for more $\Sigma$, collapses when using the scaled variables $\langle \gamma_{\rm ms}(\Sigma,L)\rangle^{-1}/|\Sigma - \Sigma_c|^\beta$ and $L^{-1}/|\Sigma - \Sigma_c|^\nu$. The collapse shown uses $\beta/\nu=0.57$, $\nu=1.8$, and $\Sigma_c = 0.111$.  }
\label{fig:gamma-ms-2D}
\end{figure}

\begin{figure}
\raggedright
(a) \\ \includegraphics[trim=0mm 0mm 0mm 0mm, clip, width=\columnwidth]{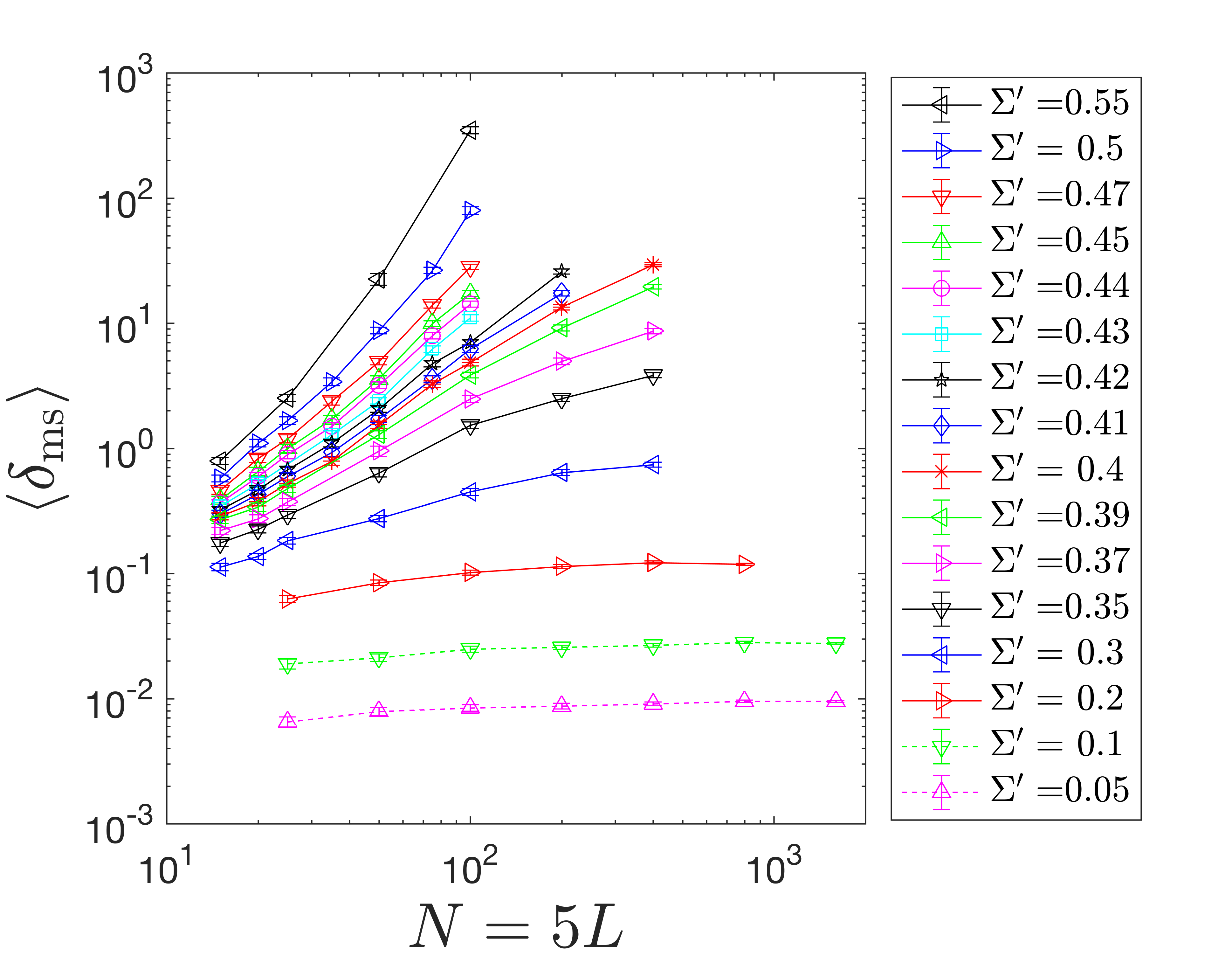} \\
(b) \\ \includegraphics[trim=0mm 0mm 0mm 0mm, clip, width=\columnwidth]{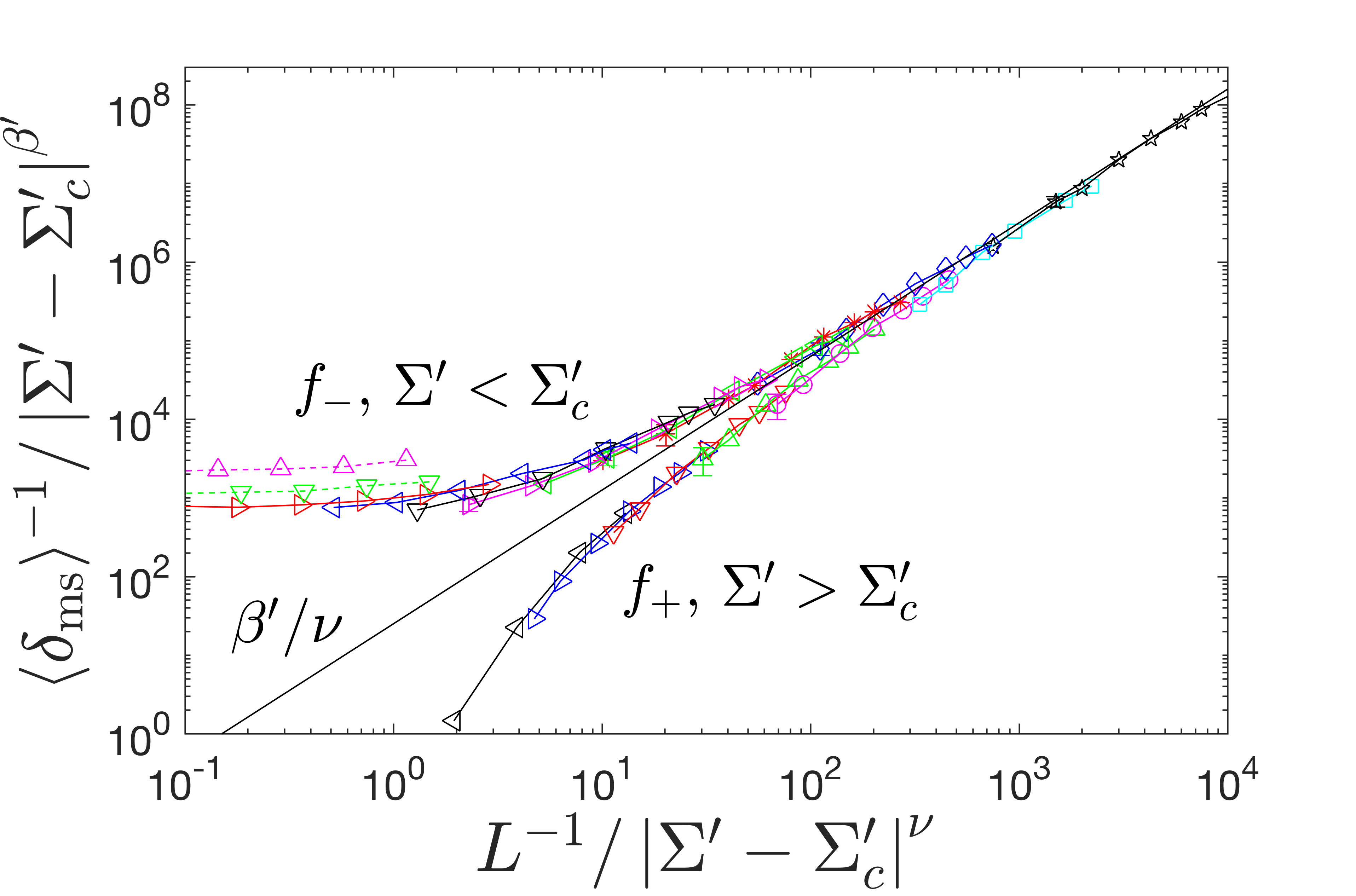} \\
\caption{The mean grain displacement $\langle \delta_{\rm ms} \rangle$ in the 2D riverbed-like geometry between initial and final MS packings plotted versus system size $N=5L$ for several values of applied stress $\Sigma'$. Solid (dashed) lines correspond to $|\Sigma'-\Sigma'_c|/\Sigma_c$ less (greater) than 0.5. Error bars are the standard error of the mean, given by the standard deviation within the sample divided by the square root of the number of trials. (b) The data from panel (a) collapses when using the scaled variables $\langle \delta_{\rm ms} (\Sigma',L) \rangle^{-1} / |\Sigma' - \Sigma'_c|^{\beta'}$ and $L^{-1}/|\Sigma' - \Sigma'_c|^\nu$. The collapse shown uses $\beta'/\nu=1.7$, $\nu=1.75$, and $\Sigma'_c = 0.41$. }
\label{fig:gamma-ms-RB}
\end{figure}

In Fig.~\ref{fig:gamma-ms-RB}, we display the results for the 2D riverbed-like geometry, which verifies that the scaling behavior is universal with respect to changes in the boundary conditions and driving method. Instead of shear strain, for each simulation we measure the average horizontal distance $\delta_{\rm ms}$ traveled by a grain between initial ($\Sigma'=0$) and final ($\Sigma'>0$) MS packings. Figure~\ref{fig:gamma-ms-RB}(a) shows the ensemble-averaged values $\langle \delta_{\rm ms}\rangle$ as a function of $\Sigma'$ and $L$. As before, these data collapse when plotted as a function of the scaled variables $L^{-1}/|\Sigma' - \Sigma'_c|^\nu$ and $\langle \delta_{\rm ms}\rangle^{-1}/|\Sigma' - \Sigma'_c|^\beta$. Using a fitting analysis similar to the one discussed above for 3D boundary-driven simple shear, we identify $\Sigma_c'= 0.41\pm 0.015$, $\beta'/\nu = 1.7 \pm 0.2$, and $\nu= 1.75 \pm 0.1$, suggesting that the scaling behavior and the value of $\nu\approx 1.7-1.8$ are generic with respect to changes in the spatial dimension, geometry, boundary conditions, and driving method. We discuss the fitting analysis for this geometry in Appendix~\ref{Appendix-crit-vals}.

\subsection{Microstructure of MS packings at varying $\Sigma$}
\label{sec:microstructure}

\begin{figure}
\raggedright
(a) \\
\includegraphics[width=\columnwidth]{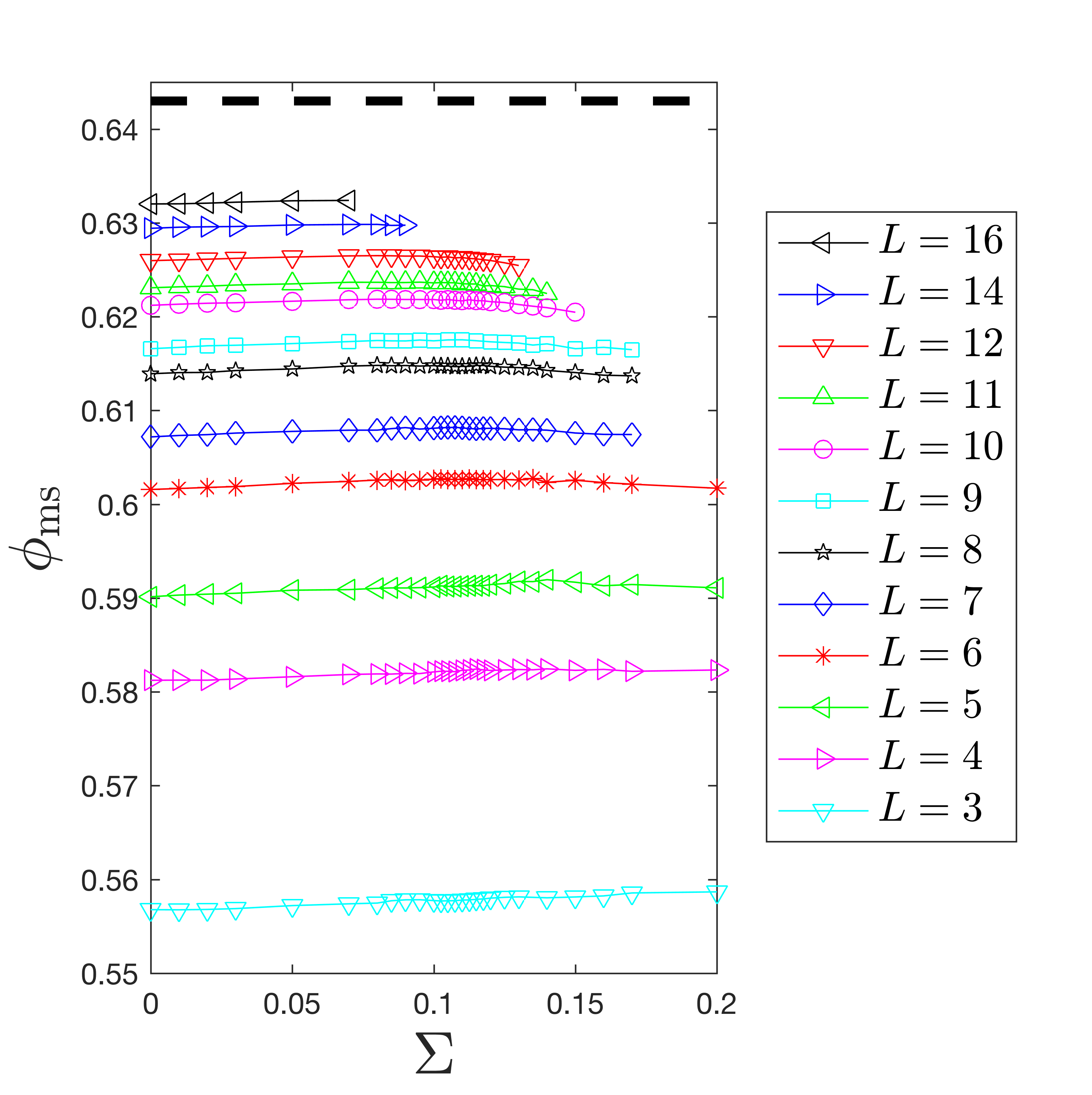} \\
(b) \\
\includegraphics[width=\columnwidth]{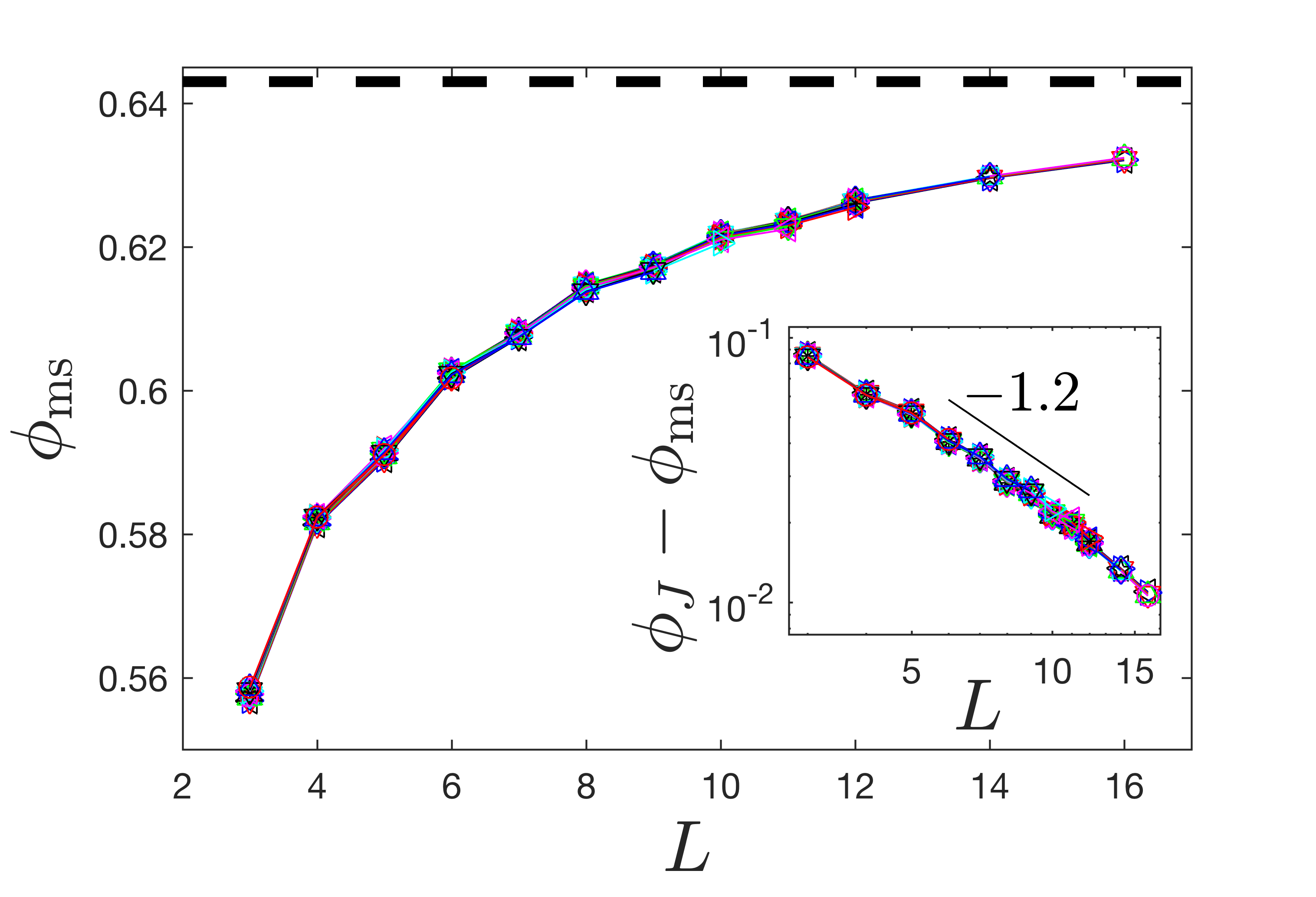}
\caption{(a) The packing fraction $\phi_{\rm ms}$ of MS packings at
jamming onset as a function of applied
shear stress $\Sigma$. Different colors represent different system
sizes. $\phi_{\rm ms}$ is
independent of $\Sigma$, and approaches $\phi_J \approx
0.643$ for large system sizes. (b) The same data from panel (a) plotted instead as a function of system size $L$. The different symbols represent different values of $\Sigma$, but the curves lie one top of one another. The inset to panel (b) shows $\phi_J-\phi_{\rm ms}$ versus $L$
plotted on a double logarithmic scale. The solid black
line has slope -1.2, implying that $\nu_J \approx 0.8$ if $(\phi_J -
\phi_{\rm ms}) \sim L^{-1/\nu}$.}
\label{fig:phi-vs-Sigma-and-N}
\end{figure}

To understand why the number density of MS packings vanishes at $\Sigma_c$, we quantify their structure using the packing fraction $\phi_{\rm ms}$ as well as the stress and contact fabric tensors. Figure~\ref{fig:phi-vs-Sigma-and-N}(a) shows a plot of packing fraction $\phi_{\rm ms}$ of MS packings generated in 3D via simple shear as a function of $\Sigma$ for varying $L$. Each data point represents the ensemble average of 200 systems. $\phi_{\rm ms}$ shows weak, nonmonotonic dependence on $\Sigma$, consistent with Fig.~10 in Ref.~\cite{peyneau08}. Specifically, $\phi_{\rm ms}$ rises slightly (by about $0.1\%$) from $\Sigma = 0$ to $\Sigma = \Sigma_c$ and then decreases slightly for $\Sigma>\Sigma_c$. Figure~\ref{fig:phi-vs-Sigma-and-N}(b) shows the same data plotted as a function of system size $L$. The different symbols represent different values of $\Sigma$, but these curves all lie on top of one another. As $L$ increases, $\phi_{\rm ms}$ approaches $\phi_J \approx 0.643$, which is indicated by a dashed black line.

The data presented in Fig.~\ref{fig:phi-vs-Sigma-and-N}(b) also allows us to estimate the critical length scale exponent $\nu_J$ for jamming. If we assume that there is a diverging length scale $\xi_J
\sim |\phi_J-\phi_{\rm ms}|^{-\nu_J}$ related to jamming that controls the system-size dependence in Fig.~\ref{fig:phi-vs-Sigma-and-N}, we expect that $L/\xi_J$ should be a constant and the packing fraction deviation scales as $(\phi_J - \phi_{\rm ms}) \sim L^{-1/\nu_J}$. The inset to Fig.~\ref{fig:phi-vs-Sigma-and-N}(a) shows that $\nu_J \approx 1/1.2 \approx 0.8$. This result is in agreement with previous studies~\cite{ohern2003,olsson2007,olsson2011}, which have estimated $\nu_J$ to be between $0.6$ and $1$. We again note that this value for $\nu_J$ is distinct from $\nu\approx 1.7-1.8$ that we estimate for yielding.

\begin{figure*}
\raggedright
(a) \hspace{39mm} (b) \hspace{39mm} (c) \hspace{39mm} (d) \\ 
\includegraphics[trim=10mm 5mm 10mm 0mm,clip,width=0.21\textwidth]{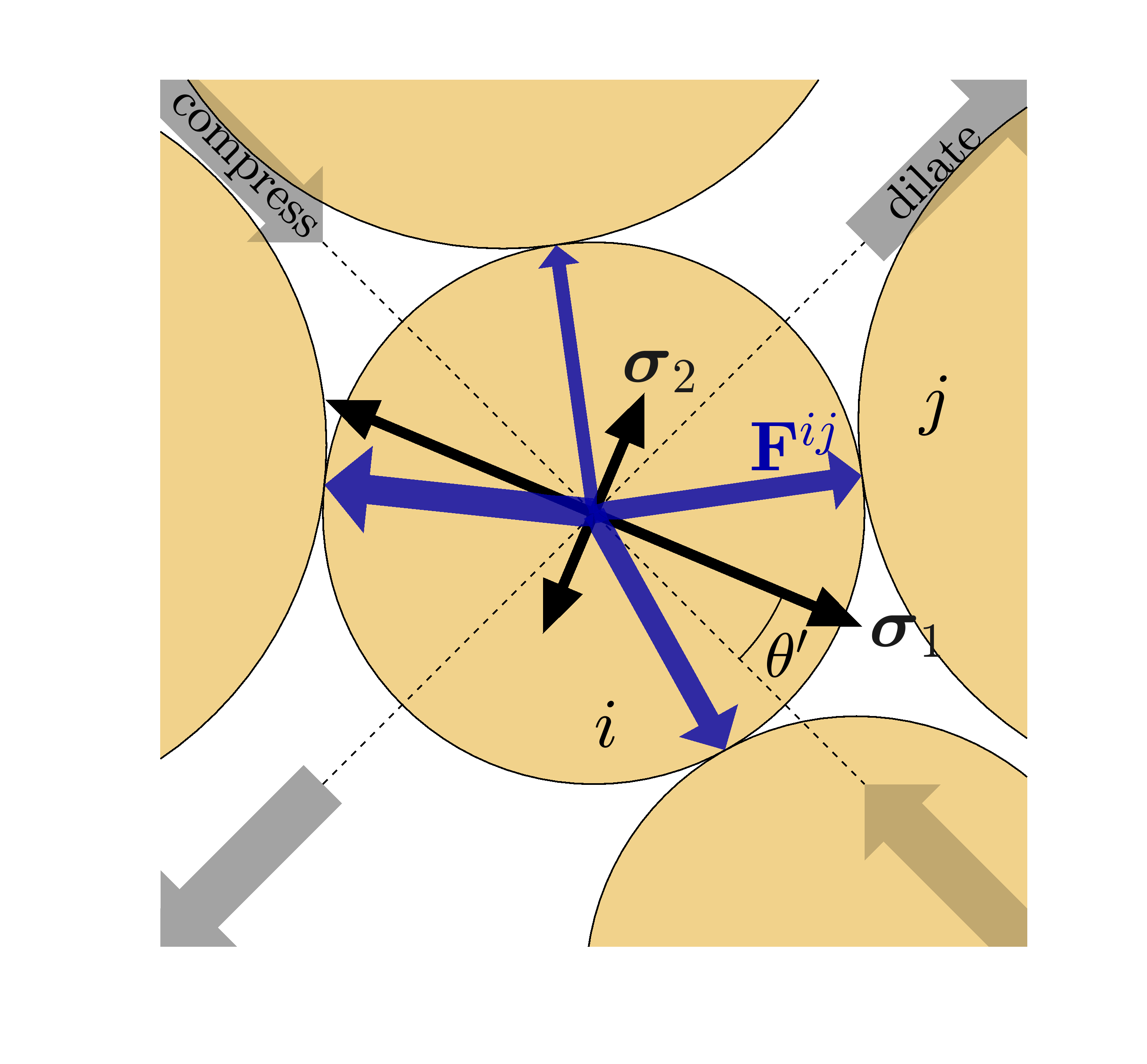}
\includegraphics[trim=0mm 0mm 0mm 2mm,clip,width=0.25\textwidth]{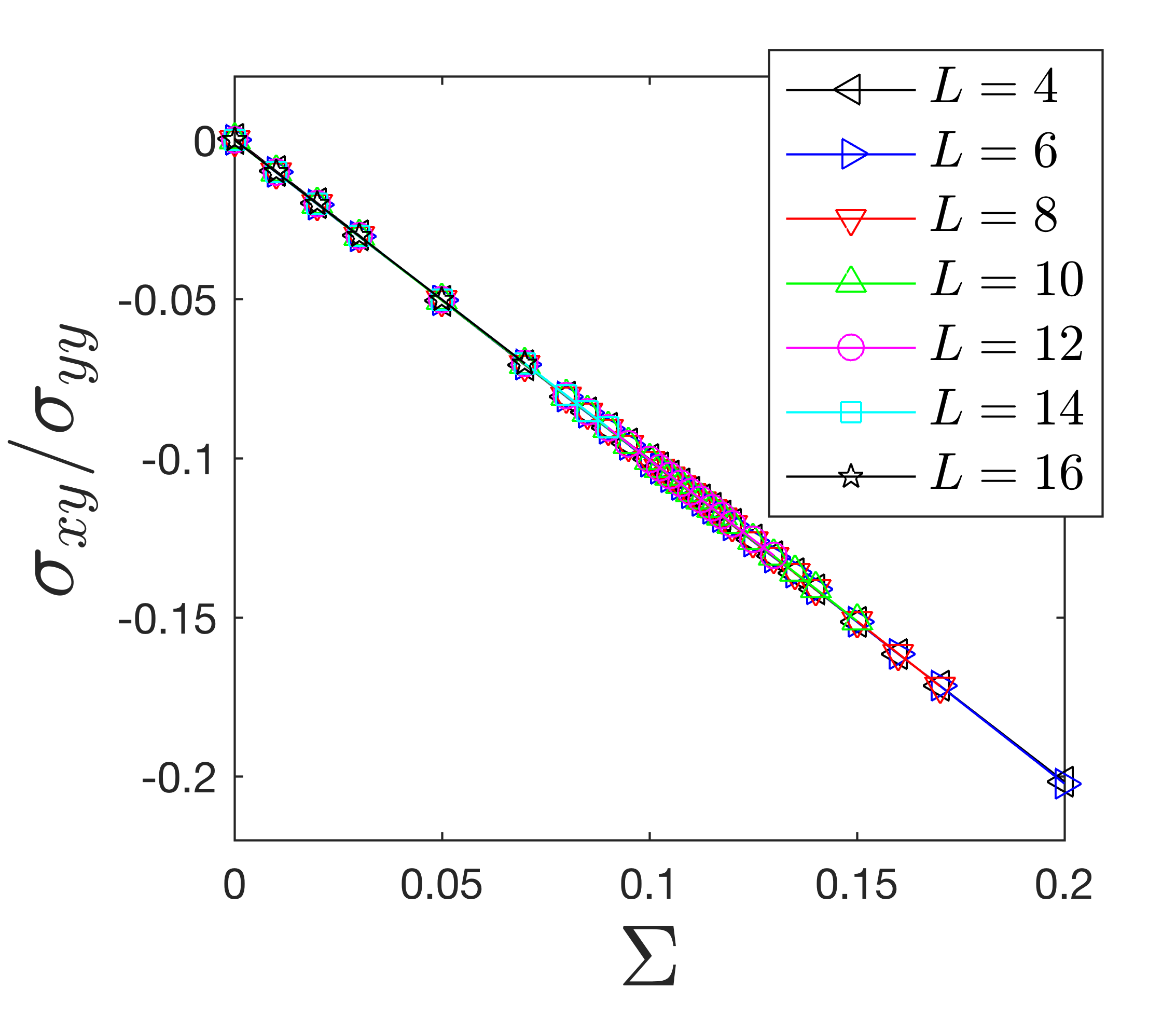}
\includegraphics[trim=0mm 0mm 0mm 2mm,clip,width=0.25\textwidth]{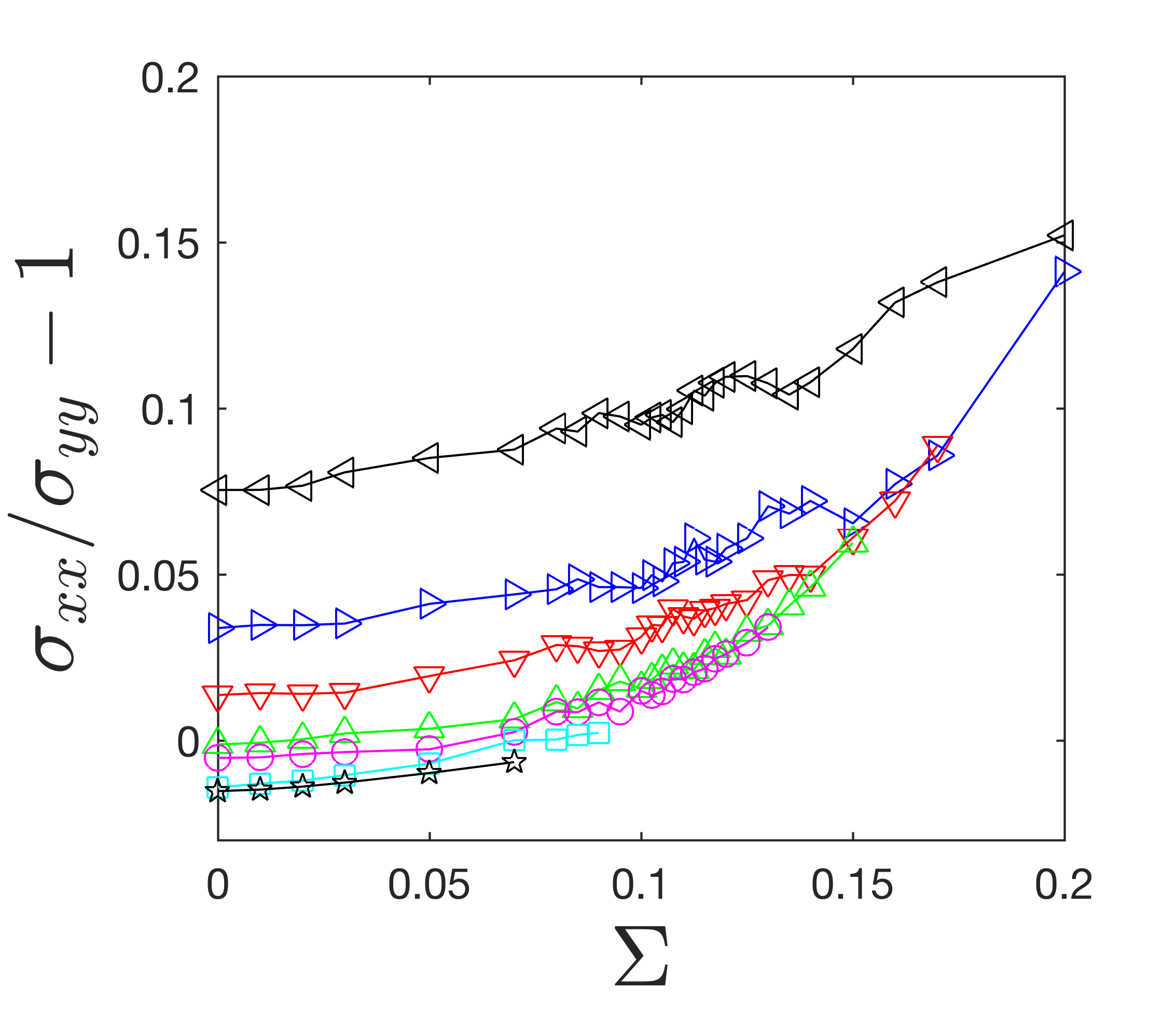}
\includegraphics[trim=0mm 0mm 0mm 2mm,clip,width=0.25\textwidth]{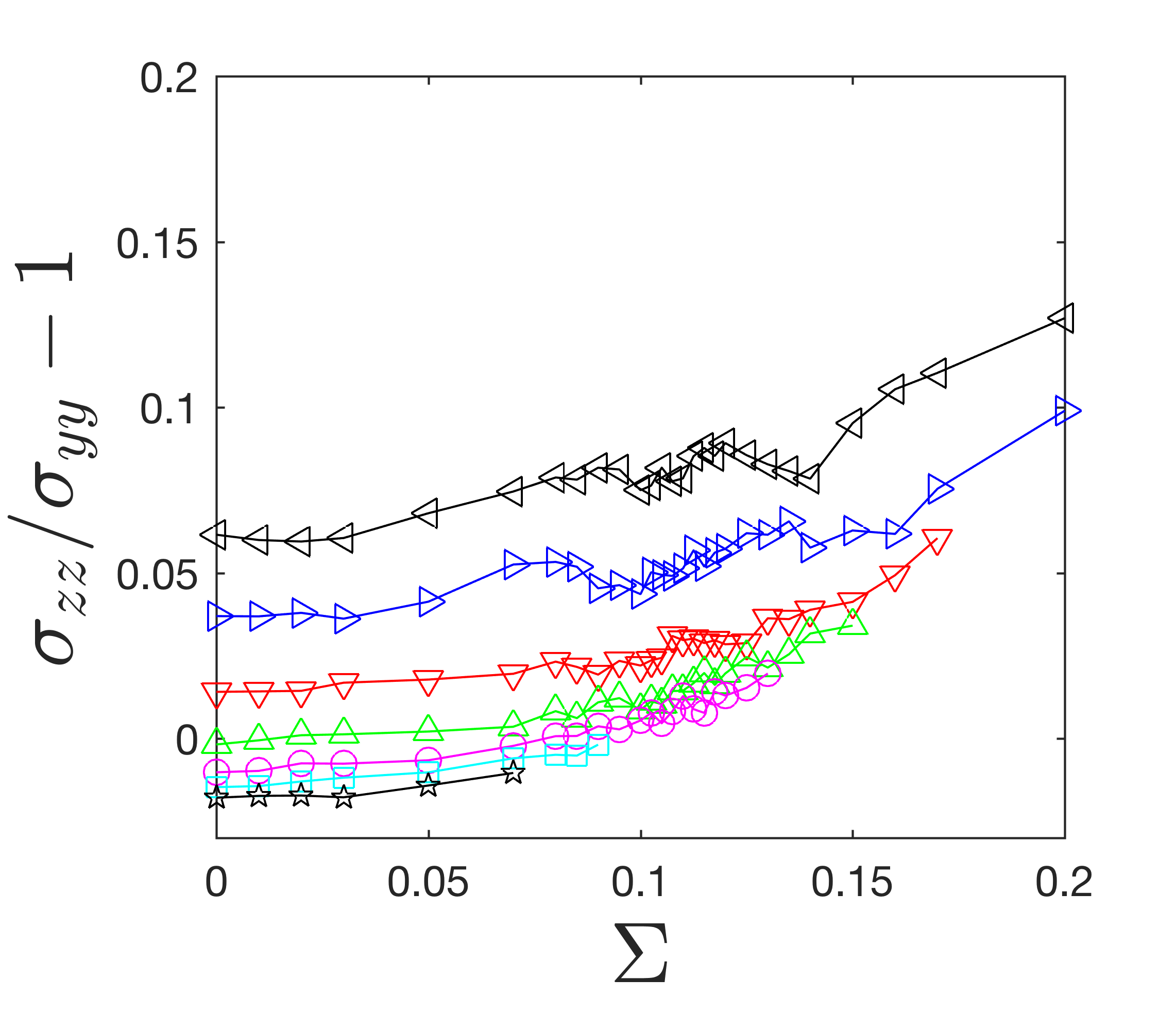}
\\(e) \hspace{39mm} (f) \hspace{39mm} (g) \hspace{39mm} (h) \\ 
\includegraphics[trim=10mm 5mm 10mm 0mm,clip,width=0.43\columnwidth]{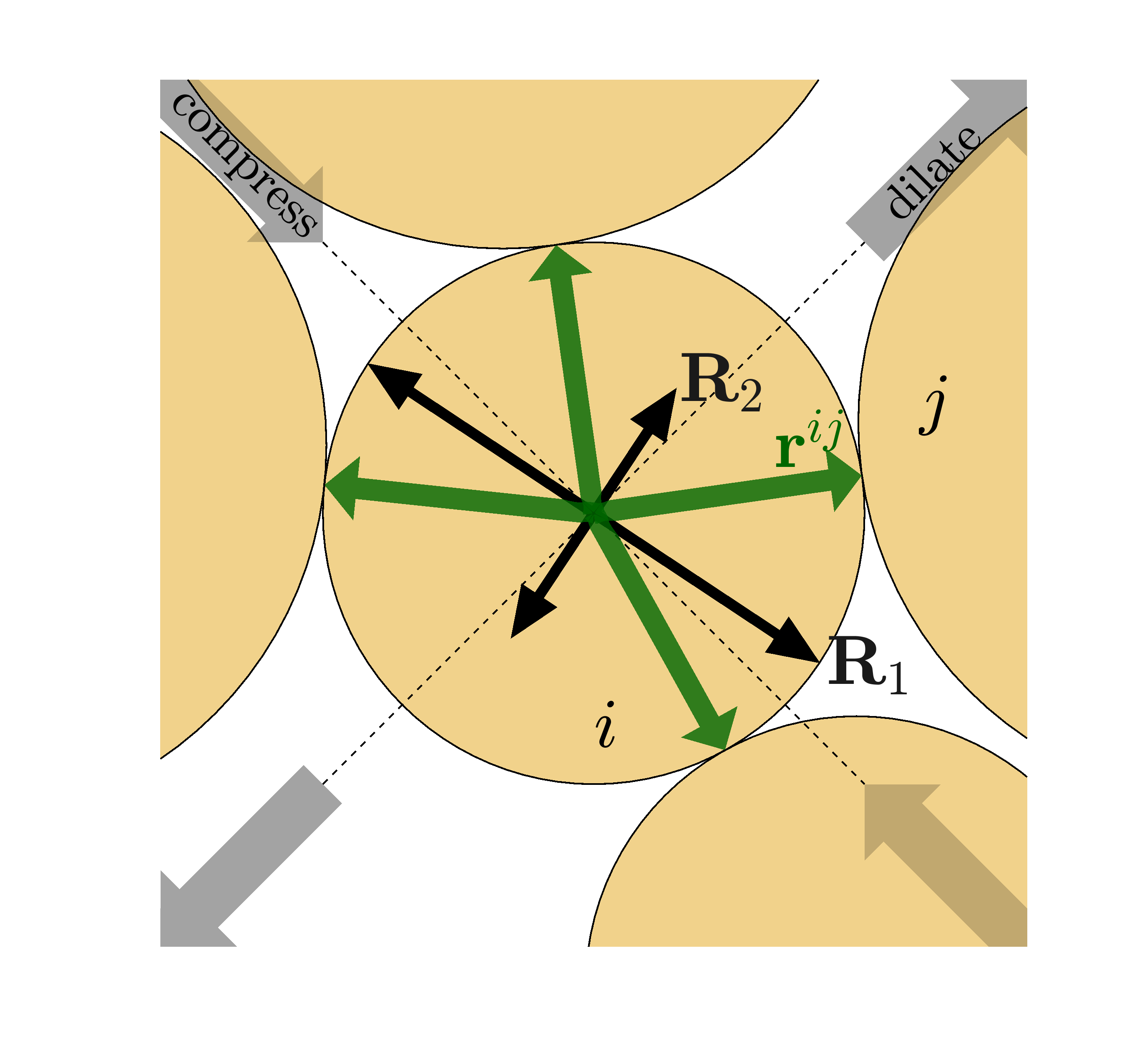}
\includegraphics[trim=0mm 0mm 0mm 2mm,clip,width=0.52\columnwidth]{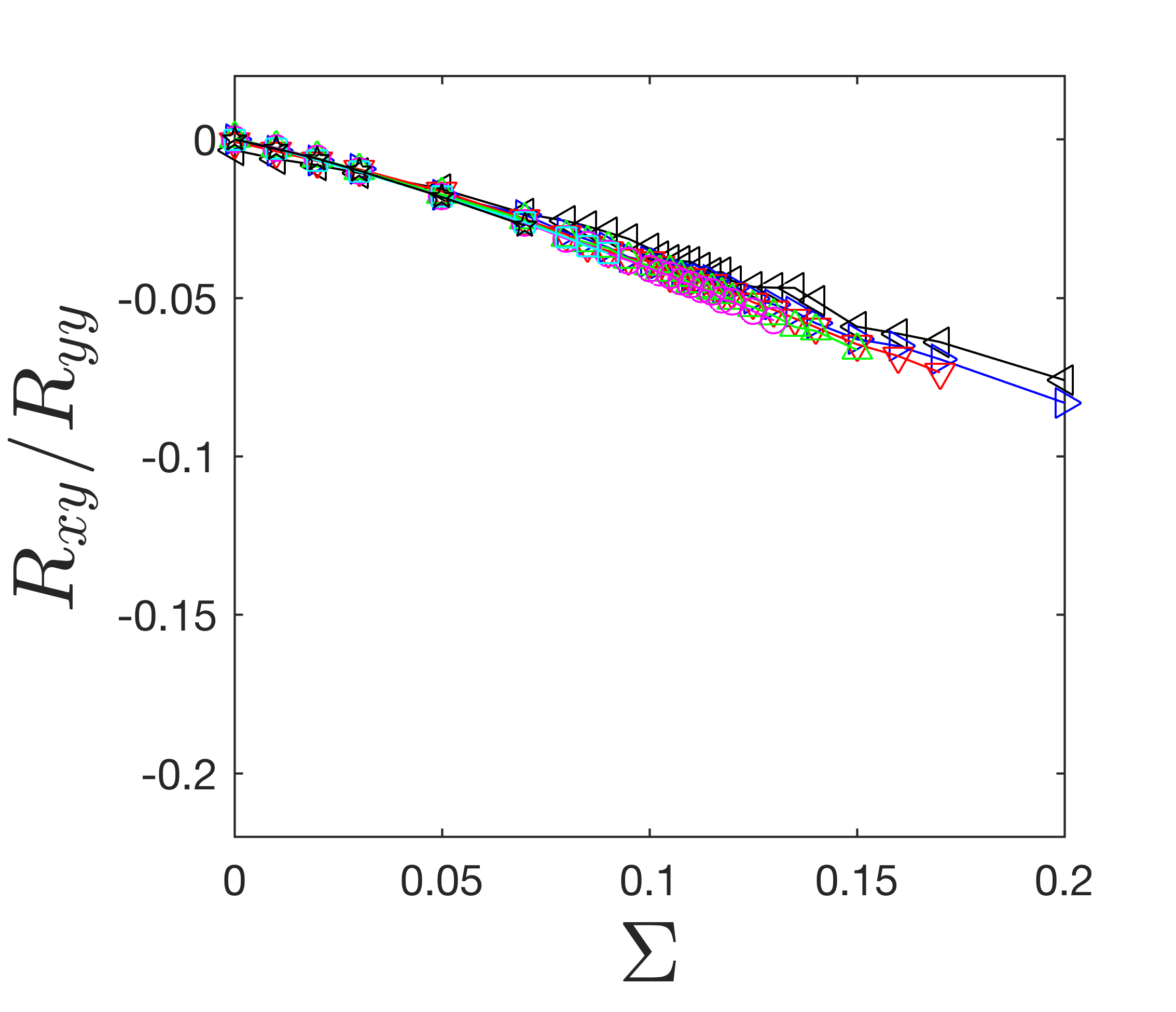}
\includegraphics[trim=0mm 0mm 0mm 2mm,clip,width=0.52\columnwidth]{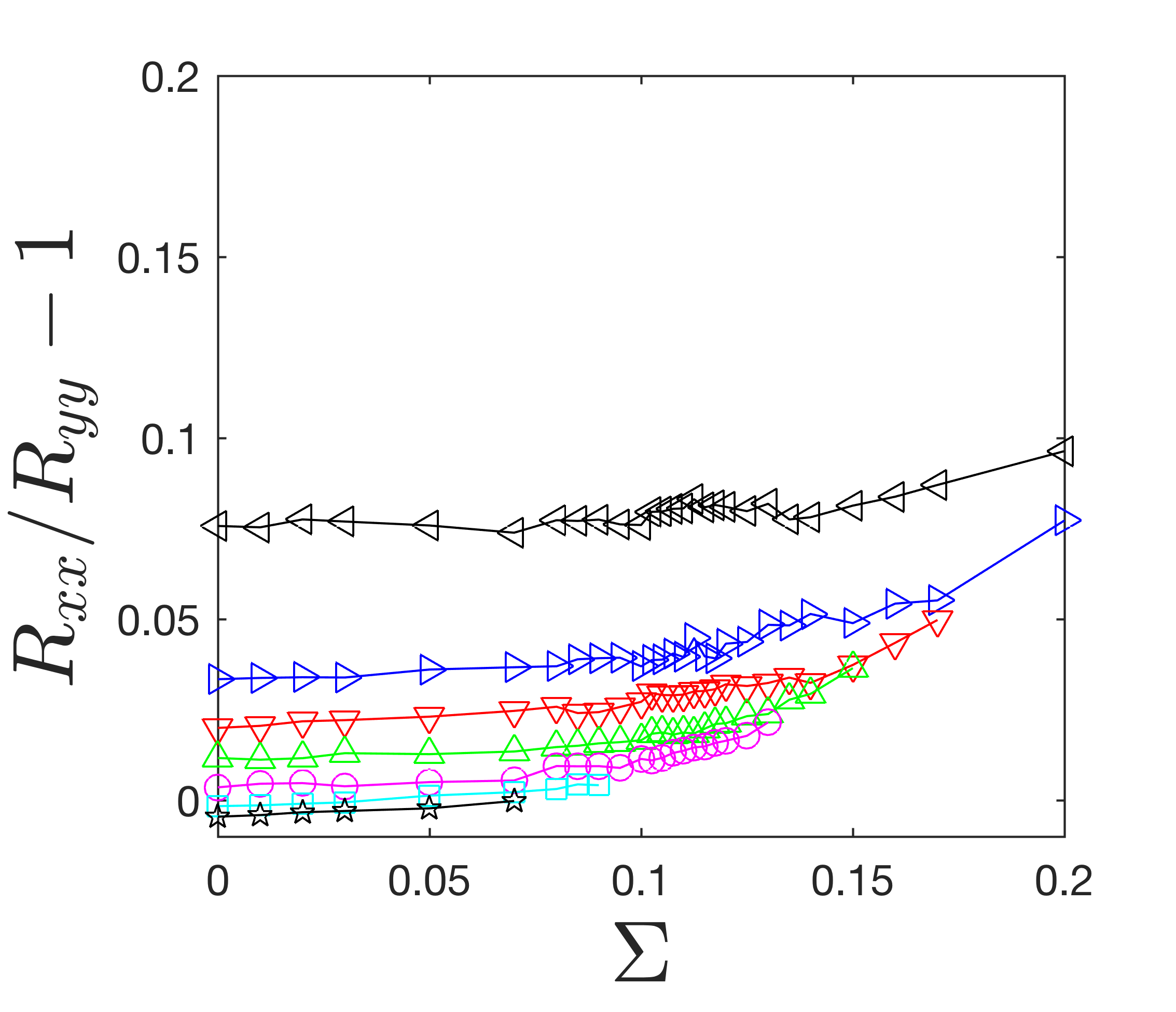}
\includegraphics[trim=0mm 0mm 0mm 2mm,clip,width=0.52\columnwidth]{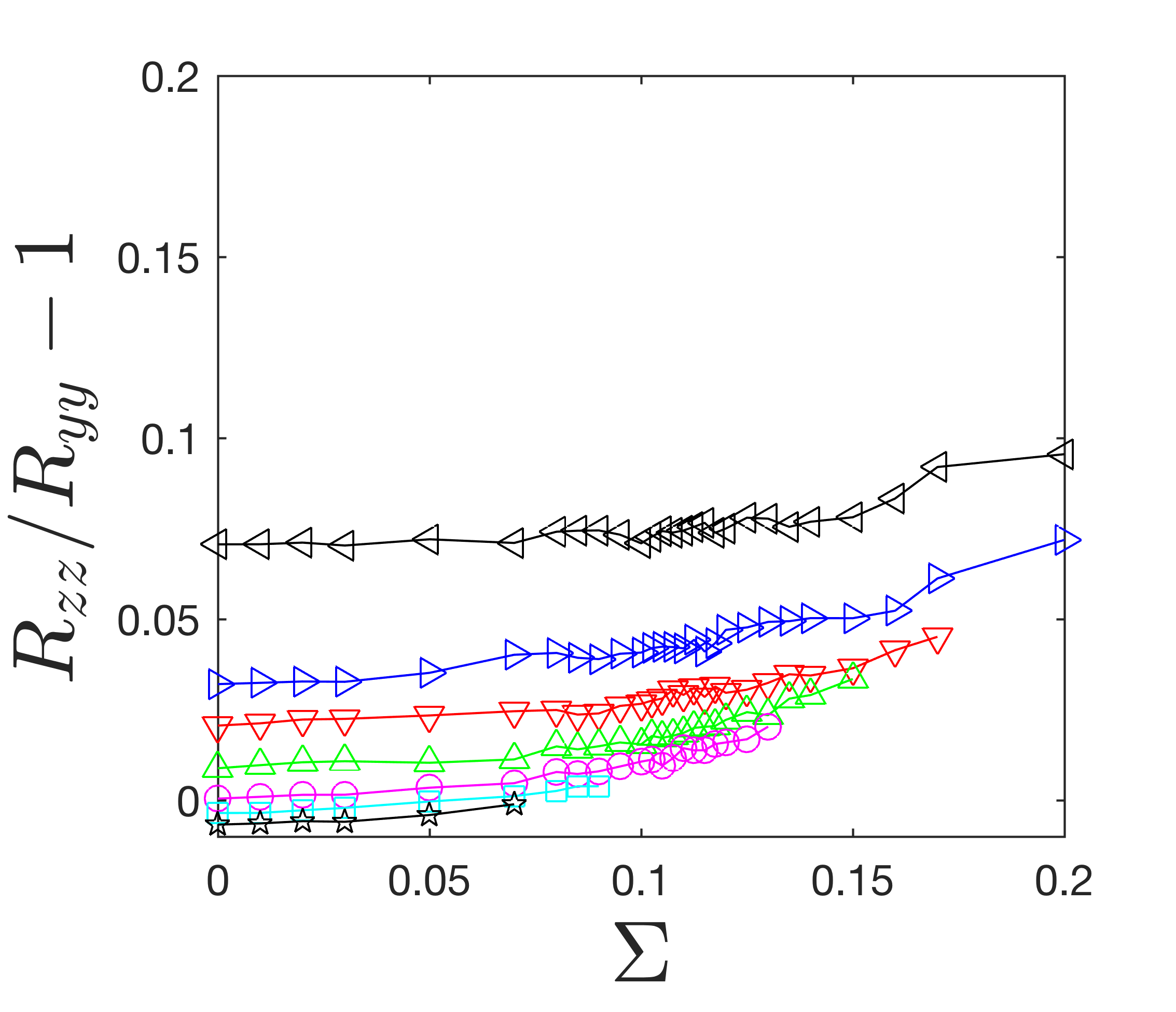}
\caption{(a,e) Close-up of MS packings in 2D illustrating features of the (a) stress and (e) fabric tensors for the central grain. ${\boldsymbol \sigma}_{1,2}$ and $\mathbf{R}_{1,2}$ denote the eigenvalue-eigenvector pairs from the sum over the contacts (blue and green arrows) for the center grain $i$ (Eqs.~\eqref{eqn:stress-tensor} and \eqref{eqn:fabric-tensor}). The magnitudes of the arrows are proportional to the eigenvalues and the directions are along the eigenvectors. $\theta'$ is the angle between the larger eigenvector and the compressive direction. (b,f) Ensemble averages of $\sigma_{xy}/\sigma_{yy}$ and $R_{xy}/R_{yy}$ for MS packings are plotted versus $\Sigma$ for varying $L$, showing $\sigma_{xy}/\sigma_{yy}=-\Sigma$ and $R_{xy}/R_{yy}\approx-0.4\Sigma$ for all $L$. (c,g) Ensemble averages of the normal anisotropies in the $x-$direction in the (c) stress tensor, $\sigma_{xx}/\sigma_{yy}-1\equiv \lambda_x$, and (g) fabric tensor, $R_{xx}/R_{yy}-1\equiv \rho_x$, are plotted versus $\Sigma$ for varying $L$. (d,h) Ensemble averages of the normal anisotropies in the $z-$direction in the (d) stress tensor, $\sigma_{zz}/\sigma_{yy}-1\equiv \lambda_z$, and (h) fabric tensor, $R_{zz}/R_{yy}-1\equiv \rho_z$, are plotted versus $\Sigma$ for varying $L$.}
\label{fig:stress}
\end{figure*}

The stress and contact fabric tensors~\cite{bi2011,baity2017} are given by
\begin{align}
\sigma_{\alpha \lambda} &= \frac{1}{V} \sum\limits_{i\neq j} r^{ij}_\alpha F^{ij}_\lambda 
\label{eqn:stress-tensor}
\\
R_{\alpha \lambda} &= \frac{1}{N} \sum\limits_{i\neq j} \frac{r^{ij}_\alpha r^{ij}_\lambda}{||\mathbf{r}^{ij}||^2}.
\label{eqn:fabric-tensor}
\end{align}
Here, $\alpha$ and $\lambda$ are Cartesian coordinates, $V$ is the system volume, $r^{ij}_\alpha$ is the $\alpha$-component of the center-to-center separation vector between grains $i$ and $j$, and $F^{ij}_\lambda$ is the $\lambda$-component of the intergrain contact force. The sum over $i$ and $j$ includes all pairs of contacting grains (excluding grain-wall contacts). 

Force balance requires $\sigma_{xy} = \sigma_{yx} = -\tau$,
$\sigma_{yy} = p$, and $\sigma_{yz} = \sigma_{zy} = 0$. In
Fig.~\ref{fig:stress}(b), we show ensemble averages of $\sigma_{xy}/\sigma_{yy}$ as a function of $\Sigma = \tau / p$. The data follows a linear relation with a slope of negative one, confirming that force balance is satisfied. We also find that $\sigma_{yz} = \sigma_{zy} = 0$ and $\sigma_{xz} = \sigma_{zx} = 0$ (not shown). Figure~\ref{fig:stress}(f) shows that the force balance criterion $\sigma_{xy}/\sigma_{yy}=-\Sigma$ requires a proportional change in the corresponding fabric tensor component, $R_{xy}/R_{yy} = - a \Sigma$ with $a \approx 0.4$. Results for 2D simple shear (not shown) are identical: we find $\sigma_{xy}/\sigma_{yy}=-\Sigma$ and $R_{xy}/R_{yy} = - a \Sigma$, but with $a \approx 0.33$. Thus, since MS packings at increasing $\Sigma$ require grain-grain contacts to be increasingly oriented along the compressive direction, the vanishing density of MS packings likely results from an upper limit of the stress and corresponding fabric anisotropies that can be realized in a large system.

Finally, we show in Fig.~\ref{fig:stress} (c), (d), (g), and (h) the excess normal stresses $\sigma_{xx}/\sigma_{yy} - 1 \equiv \lambda_x$ and $\sigma_{zz}/\sigma_{yy} - 1 \equiv \lambda_z$ as well as the corresponding quantites from the fabric tensor $R_{xx}/R_{yy} - 1 \equiv \rho_x$ and $R_{zz}/R_{yy} - 1 \equiv \rho_z$. These quantities represent excess compressive stresses and contacts that exist in the periodic $x-$ and $z-$directions. For $\Sigma<\Sigma_c$, $\lambda_{x,z}$ and $\rho_{x,z}$ begin at some finite value and tend to zero at large $L$. For $\Sigma>\Sigma_c$, $\lambda_{x,z}$ and $\rho_{x,z}$ increase with $\Sigma$. We find similar results for 2D simple shear (not shown).

To understand why the normal stress and fabric anisotropies increase with $\Sigma$, we consider the ensemble-averaged stress tensor $\langle \boldsymbol{\sigma} \rangle$ of MS packings in 3D at a given $\Sigma$, which can be written as
\begin{equation}
\langle \boldsymbol{\sigma} \rangle= p \begin{bmatrix}
    1+\lambda_x       & -\Sigma  & 0 \\
   -\Sigma       & 1 & 0 \\
   0 & 0 & 1+\lambda_z
\end{bmatrix}.
\label{eq:stress-tensor-MS}
\end{equation}
We consider only the stress components in the $x$-$y$ plane, which are
decoupled from $z$ in Eq.~\eqref{eq:stress-tensor-MS}, and its
eigenvalue-eigenvector pairs $\{\sigma_1,{\boldsymbol \sigma}_1\}$ and
$\{\sigma_2,{\boldsymbol \sigma}_2\}$. The internal stress anisotropy
is $\Sigma_i = \tau_i / p_i$, where $p_i = (\sigma_1 + \sigma_2)/2$
and $\tau_i = (\sigma_1 - \sigma_2)/2$ are the internal pressure and
shear stress, respectively. From Eq.~\eqref{eq:stress-tensor-MS}, $\Sigma_i = \frac{\sqrt{4\Sigma^2 +
    \lambda_x^2}}{2+\lambda_x}$, and ${\boldsymbol \sigma}_1$ is
oriented at an angle that deviates from the compression direction by
an angle $\theta'$, as shown in Fig.~\ref{fig:stress}(a). An
expansion of $\Sigma_i$ for small $\lambda_x$ gives $\theta' =\frac{
  \lambda_x}{4 \Sigma} +
\mathcal{O}\left[\left(\frac{\lambda_x}{\Sigma}\right)^3\right]$. 

Thus, when $\lambda_x = 0$, $\Sigma_i = \Sigma$ and ${\boldsymbol\sigma}_1$
and ${\boldsymbol\sigma}_2$ are aligned with the compression and
dilation directions, respectively. However, $\Sigma_i$ is minimized by
a positive, nonzero value of $\lambda_x = 2\Sigma^2$ with
$\Sigma_i^{\rm min} = \Sigma/\sqrt{1+\Sigma^2}$. This can give
$\Sigma_i<\Sigma$, but this rotates the larger eigenvector
${\boldsymbol \sigma}_{1}$ away from the compression direction.

Near $\Sigma_c$ for finite systems, MS packings
are scarce, and $\Sigma_i<\Sigma$ with $\theta'>0$ may be
preferable, despite the broken symmetry. However, the broken symmetry becomes
more difficult to achieve for larger systems. We note that the
dependence of $\lambda_x$, $\lambda_z$, $\rho_x$, and $\rho_z$ on $L$ in
Fig.~\ref{fig:stress}(c,d) and (g,h) is suggestive of critical scaling
(which we expect if $\xi$ dominates the behavior near $\Sigma_c$)
similar to Eq.~\eqref{eq:scaling-1}. The scaling results for these quantities are not as conclusive, and we leave a more extensive study of the possible scaling of these quantities for future work. 


\section{Conclusion}
\label{sec:conclusion}

In conclusion, for frictionless spherical grains under shear, we find that the number of MS packings vanishes near $\Sigma=\Sigma_c \approx 0.11$. Finite-size effects depend on a diverging length scale $\xi \propto |\Sigma - \Sigma_c|^{-\nu}$. We find similar results for the cases of 3D simple shear, shown in Fig.~\ref{fig:gamma-ms-3D}, for 2D simple shear, shown in Fig.~\ref{fig:gamma-ms-2D}, and in a 2D riverbed-like geometry, shown in Fig.~\ref{fig:gamma-ms-RB}. Thus, the critical scaling behavior, including the value of the exponent $\nu\approx 1.7-1.8$, is generic with respect to changes in spatial dimension, system geometry, and boundary conditions. 

We find that the packing fraction of MS packings at varying $\Sigma$ shows weak, nonmonotonic dependence on $\Sigma$, in agreement with previous work~\cite{peyneau08}. This suggests that the critical scaling we observe is distinct from that associated with jamming. The force balance criterion, Fig.~\ref{fig:stress}(b), is accompanied by a proportional change in the fabric tensor, Fig.~\ref{fig:stress}(e). Thus, we argue that $\Sigma_c$ corresponds to the maximum anisotropy that can be realized in the large-system limit. This hypothesis is consistent with our finding that finite-sized MS packings with $\Sigma$ near or above $\Sigma_c$ tend to be rotated relative to the axes of the applied deformation, which can reduce the internal force anisotropy of MS packings. However, this effect appears to vanish in the large-system limit, where symmetry dictates that compressive direction be aligned with the largest eigenvalues of the stress and fabric tensors for MS packings.

Finally, we note recent work on jamming by shear~\cite{bi2011,bertrand2016,baity2017,chen2018}, where MS packings obtained via simple or pure shear at constant volume also display anisotropic stress and contact fabric tensors. These results are distinct from those presented here, since we control normal stress and allow volume to fluctuate. However, we expect future work to unify these two approaches, providing a complete theory of the density of MS packings as a function of volume, stress state, preparation history, and friction.

\appendix

\section{Equations of motion}
\label{Appendix-EOM}

\subsection{Boundary-driven simple shear in 2D and 3D}
\label{Appendix-EOM-SS}
For the simple shear simulations in $d=2$ (2D) and $d=3$ (3D) spatial dimensions, we solve
Newton's equations of motion for all bulk grains as well as the top plate. The
equation of motion for the top plate is
\begin{equation}
M\mathbf{a}_{\rm plate} = \sum_{i,j} \mathbf{F}_{ij}^c+\mathbf{F}_{\rm ext}-B_{\rm plate}\mathbf{v}_{\rm plate},
\label{eq: Wall Motion}
\end{equation}
where $M$ is the plate mass, $\mathbf{a}_{\rm plate}$ is the plate acceleration, $\mathbf{F}_{ij}^c$ is the contact force on plate particle $i$ due to bulk grain $j$, $\mathbf{F}_{ext}$ is the external force exerted on the top wall, $B_{\rm plate}$ is a viscous drag coefficient, and $\mathbf{v}_{\rm plate}$ is the top plate velocity. Similarly, the equation of motion for each bulk grain $i$ is given by
\begin{equation}
m_i\mathbf{a}_i = \sum_j \mathbf{F}_{ij}^c-B_i\mathbf{v}_i,
\label{eq: Grain Motion}
\end{equation}
where $m_i \propto (D_i)^{d}$ is the mass of grain $i$ ($D_i$ is the
diameter of grain $i$), $\mathbf{a}_i$ is the acceleration of bulk
grain $i$, $\mathbf{F}_{ij}^c$ is the contact force on bulk grain
$i$ due to bulk grain $j$, $B_i$ is the drag coefficient,
and $\mathbf{v}_i$ is the velocity of bulk grain $i$. For pairwise contact forces between two bulk grains or between a bulk grain and a plate particle, we use
\begin{equation}
\mathbf{F}_{ij}^c = K\left(\frac{D_{ij}}{r_{ij}}-1\right)\Theta\left(1-\frac{r_{ij}}{D_{ij}}\right)\hat{r}_{ij},
\label{eq: Spring Force}
\end{equation}
where $K$ is a force scale, $r_{ij}$ is the center-to-center distance
between grains $i$ and $j$, $D_{ij}$ is the average diameter of
grains $i$ and $j$, $\Theta$ is the Heaviside step function, and
$\hat{r}_{ij}$ is the unit vector from the center of grain $i$ to
the center of grain $j$.

The external force on the wall is given by 
\begin{equation}
\mathbf{F}_{\rm ext} = (\tau \mathbf{\hat{x}} - p \mathbf{\hat{y}})(LD)^{d-1},
\label{eq: PTau}
\end{equation}
where $\tau$ and $p$ are the shear stress and normal stress,
respectively. Each plate particle has the same mass $m$ and drag
coefficient $B$ as the small bulk grains, and thus $M = m L^{d-1}$
and $B_{\rm wall} = BL^{d-1}$. Since the number of contacts also
scales as $L^{d-1}$, all quantities in Eq.~\eqref{eq: Wall Motion}
scale as $L^{d-1}$. Equations \eqref{eq: Wall Motion} and \eqref{eq: Grain Motion} are then governed by the nondimensional parameters given in Eq.~\eqref{eq:gamma-kappa-sigma}.

\subsection{Riverbed model in 2D}
\label{Appendix-EOM-RB}

The riverbed-like geometry that we study consists of a 2D domain of width $L$, with periodic boundary conditions horizontally, containing $N/2$ large and $N/2$ small disk-shaped grains with diameter ratio $r=1.4$. We use $N = 5L$ grains, so the beds have a height $H \approx 5D$, where $D$ is the diameter of a small grain. There is no upper boundary, and the lower boundary is rigid with a no-slip condition for any grain contacting it. The net force on each grain is given by the sum of contact forces from all other grains, a gravitational force, and a Stokes-drag-like force from a fluid flow that increases linearly with height and moves purely horizontally:
\begin{equation}
m_i\mathbf{a}_i=\sum_j \mathbf{F}^c_{ij} - m_ig'\hat{y} + B_i\left(v_0 \frac{y_i}{H}\hat{x}-\mathbf{v}_i \right).
\label{eqn:force-law}
\end{equation} 
Here, $m_i\propto D_i^2$ is the grain mass, $\mathbf{v}_i$ and $\mathbf{a}_i$ are the velocity and acceleration, respectively, of each grain $i$, $m_ig'$ is the buoyancy-corrected grain weight, $B_i$ is the drag coefficient on grain $i$, $v_0$ is a characteristic fluid velocity at the top of the bed ($y=H$), and $y_i$ is the height above the lower boundary of the center of grain $i$. The contact force $\mathbf{F}^c_{ij}= K\left(1-\frac{r_{ij}}{D_{ij}}\right)\theta\left(1-\frac{r_{ij}}{D_{ij}}\right)\hat{r}_{ij}$ is identical to the one discussed above for simple shear. Equation~\eqref{eqn:force-law} is governed by the three nondimensional parameters given in Eq.~\eqref{eqn:nondimen-params}.

\section{Determining critical values}
\label{Appendix-crit-vals}

\begin{figure}
\raggedright (a) \hspace{39mm} (b) \\ \centering
\includegraphics[width=0.49\columnwidth]{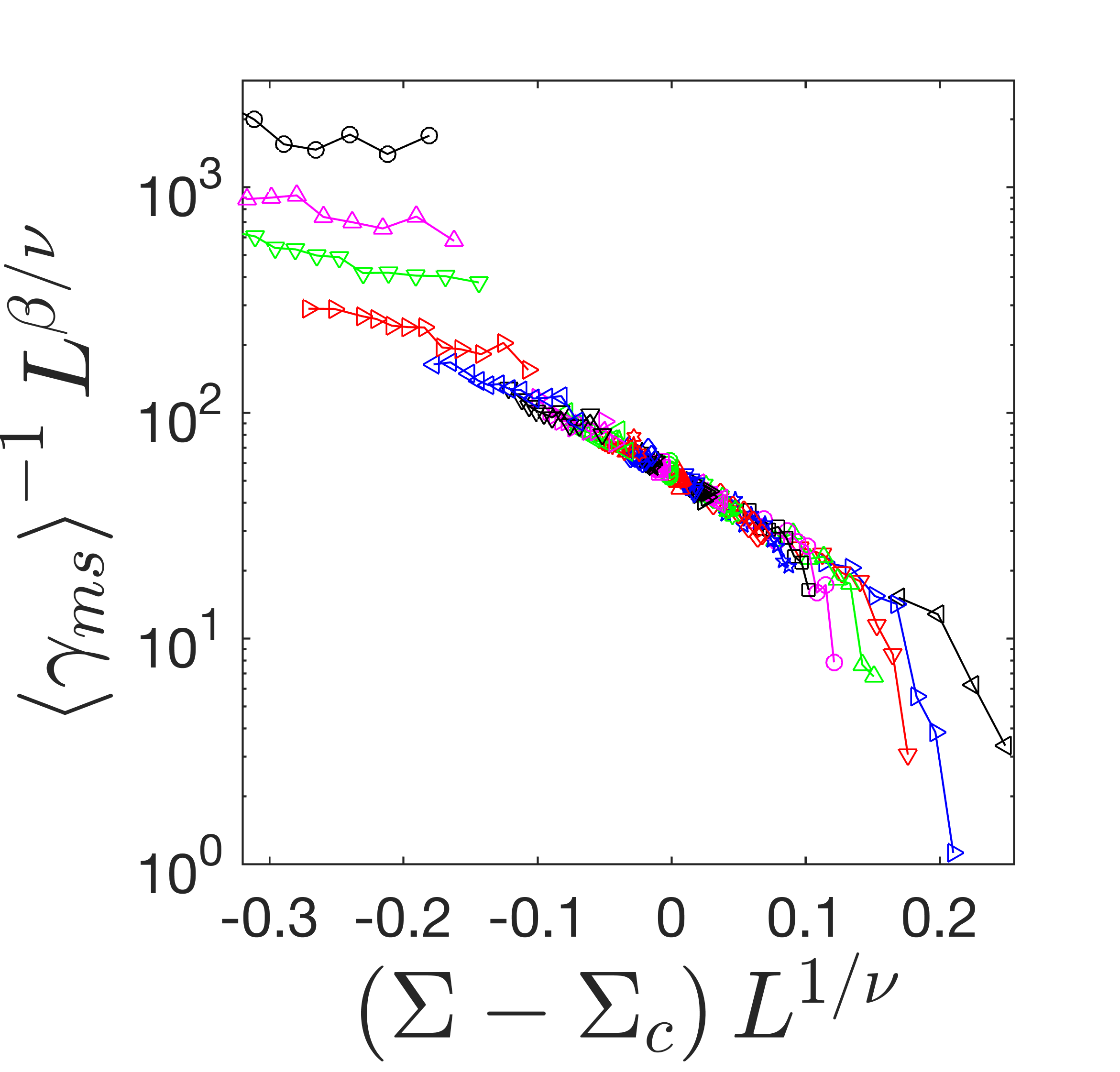}
\includegraphics[width=0.49\columnwidth]{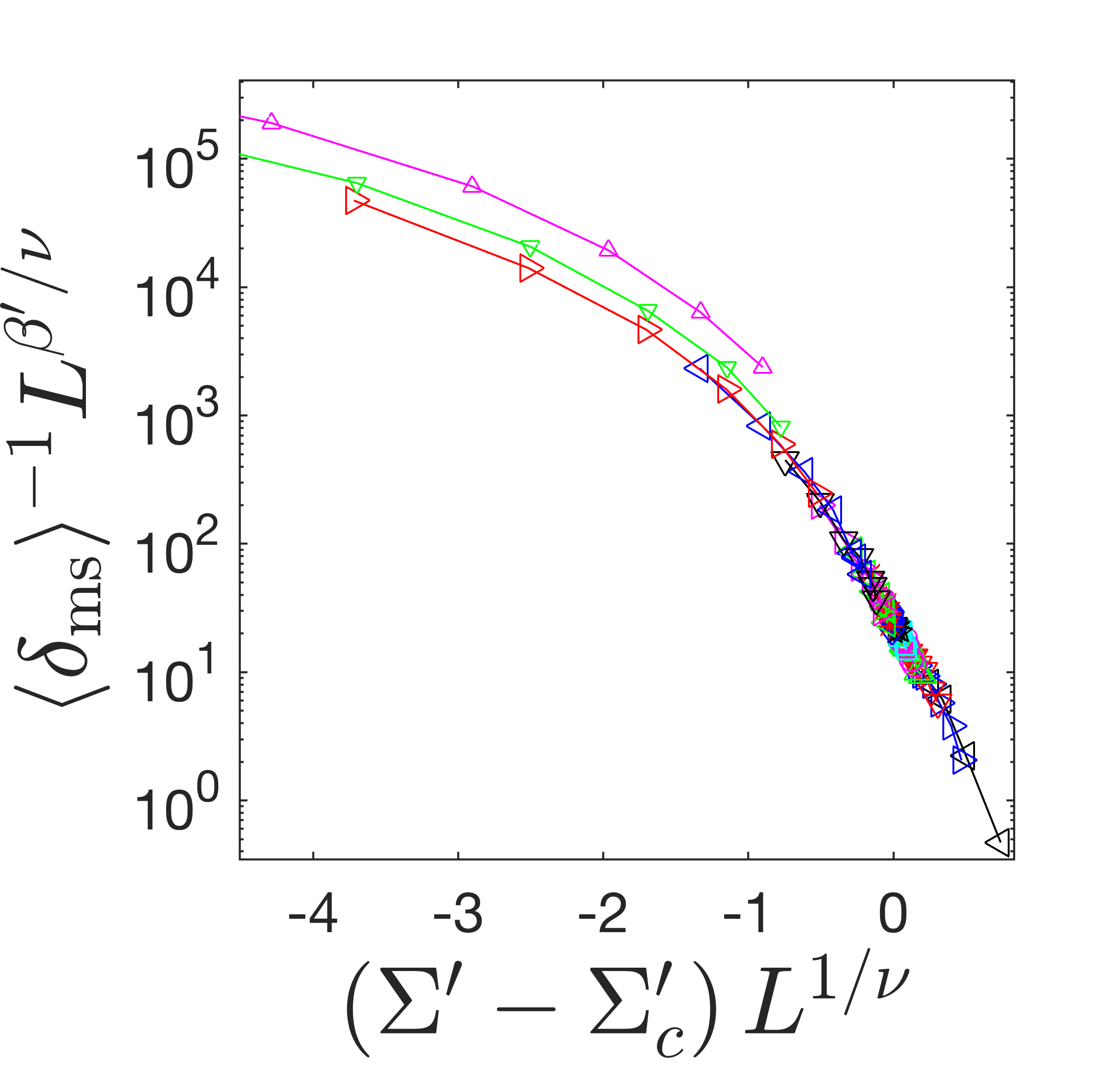}
\\ \raggedright
\caption{Data for the scaled $\langle \gamma_{\rm ms} \rangle^{-1}$ versus scaled $(\Sigma-\Sigma_c)$ using the scaling function in  Eq.~\eqref{eqn:alternate-scaling} for (a) 3D simple shear and (b) 2D riverbed geometries.}
\label{fig:alternate-scaling}
\end{figure}

As discussed above, we determine the critical exponents $\nu$
and $\beta$ as well as the value of the yield stress $\Sigma_c$ by
fitting scaled data to a scaling function with the values of
$\nu$, $\beta$, and $\Sigma_c$ treated as fit parameters. We
first estimate $\nu$, $\beta$, and $\Sigma_c$ by collapsing
the data according to 
\begin{equation}
\langle \gamma_{\rm ms} \rangle^{-1} = L^{-\beta/\nu}g \left( (\Sigma-\Sigma_c)L^{-1/\nu} \right),
\label{eqn:alternate-scaling}
\end{equation}
as shown in Fig.~\ref{fig:alternate-scaling}. This form is equivalent
to Eq. (1) in the main text, but it is more convenient to use since the
scaling function $g$ has only one branch. We fit
$\log(\langle\gamma_{\rm ms} \rangle ^{-1} L^{\beta/\nu})$ and
$(\Sigma-\Sigma_c)L^{-1/\nu}$ to a third-order polynomial. The
polynomial coefficients returned from this fit are then used as the
initial values in a Levenberg-Marquardt fit to the scaling form in
Eq. (1) in the main text, where the critical values $\Sigma_c$, $\nu$,
and $\beta$ are then used as fit parameters.

From Fig.~\ref{fig:alternate-scaling}, it is obvious that the data for
large deviations $|\Sigma-\Sigma_c|$ does not collapse as well as the
data for small deviations. In addition, we expect that data for small
system sizes does not obey the scaling collapse. Thus, we
systematically vary the range $X\equiv |\Sigma-\Sigma_c|/\Sigma_c < X_{\rm max}$ and
the minimum system size $L_{\rm min}$ that we include in our fits, although we
are somewhat limited in the maximum $L_{\rm min}$ we can use before we
no longer have enough data for a meaningful fit. We quantify the fits using the reduced chi-squared metric, $\chi^2 = \sum_i (\Delta_i)^2/e_i^2$, where the sum is over all data points $i$ used in the fit (a subset of those shown in Fig.~\ref{fig:alternate-scaling}), $\Delta_i$ is the difference between the data and the fit, and $e_i$ is the standard error of the mean, which we estimate by the standard deviation within that sample divided by the square root of the number of trials. We then measure $\chi^2/n$, where $n$ is the number of data points minus the number of fit parameters in the model. A good fit is characterized by a value of $\chi^2/n \approx 1$.

\begin{figure}[t!!]
\raggedright
(a) \hspace{37mm} (b)  \\ 
\includegraphics[width=0.48\columnwidth]{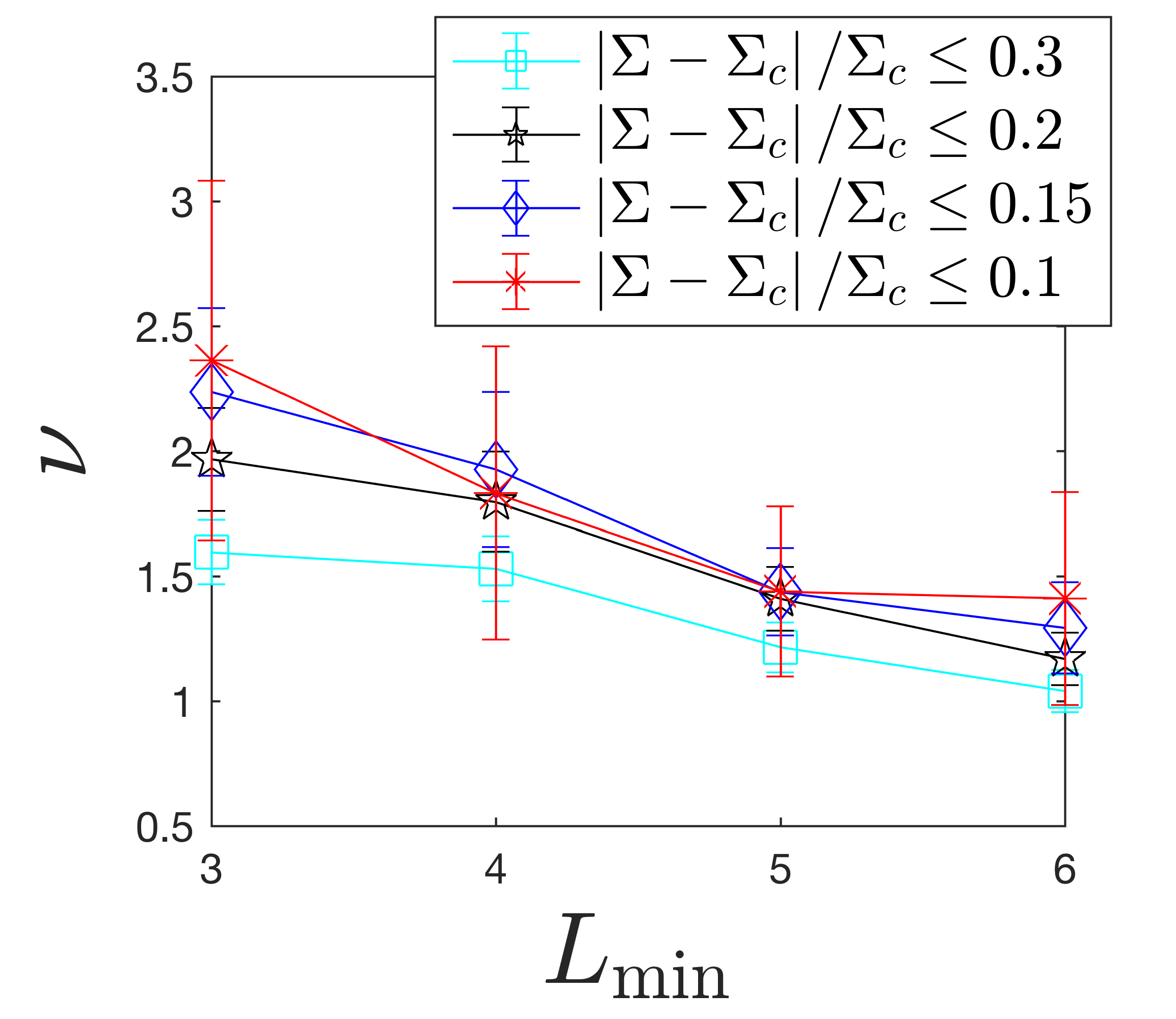}
\includegraphics[width=0.48\columnwidth]{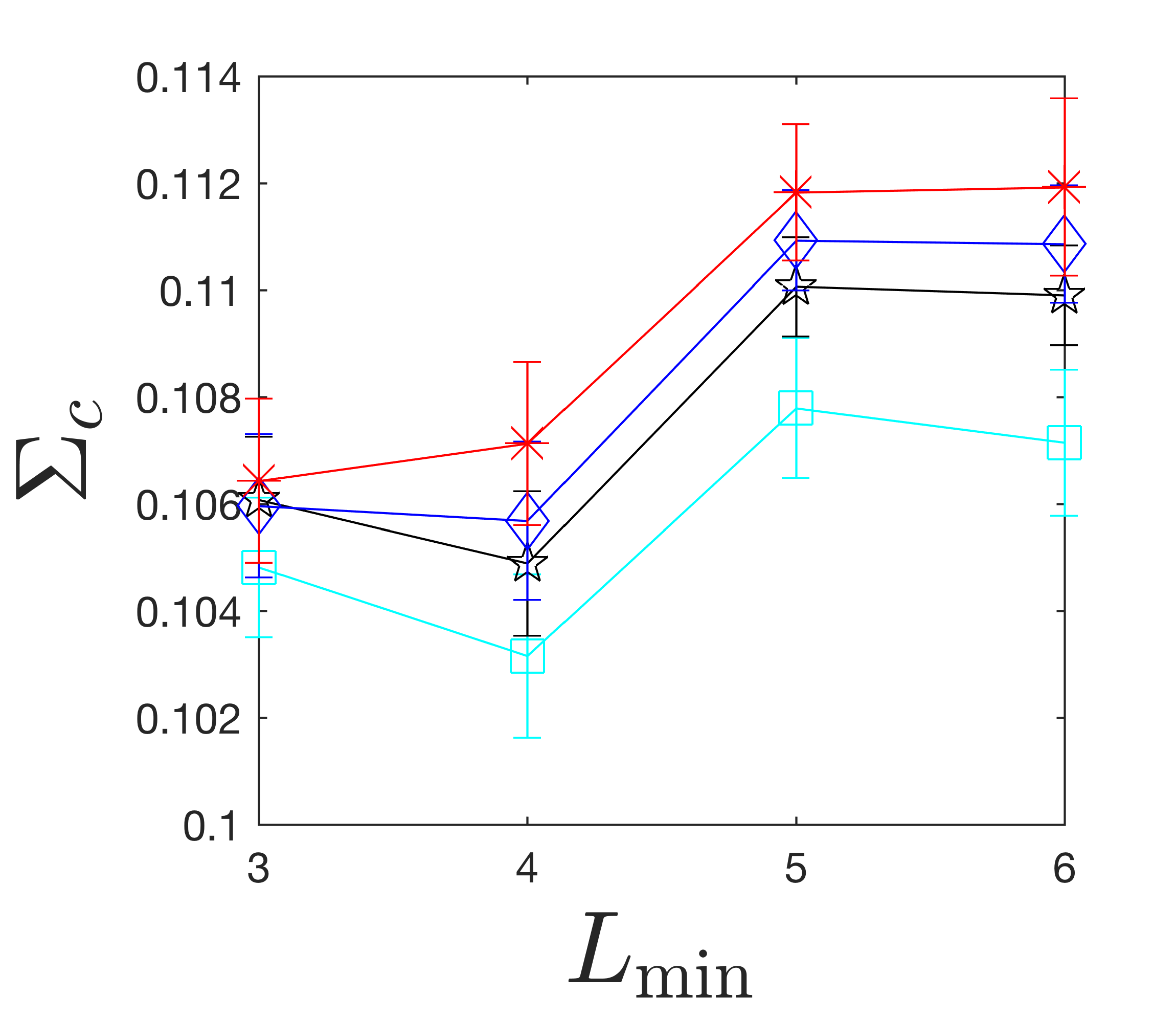}
\\ (c) \hspace{37mm} (d)  \\ 
\includegraphics[width=0.48\columnwidth]{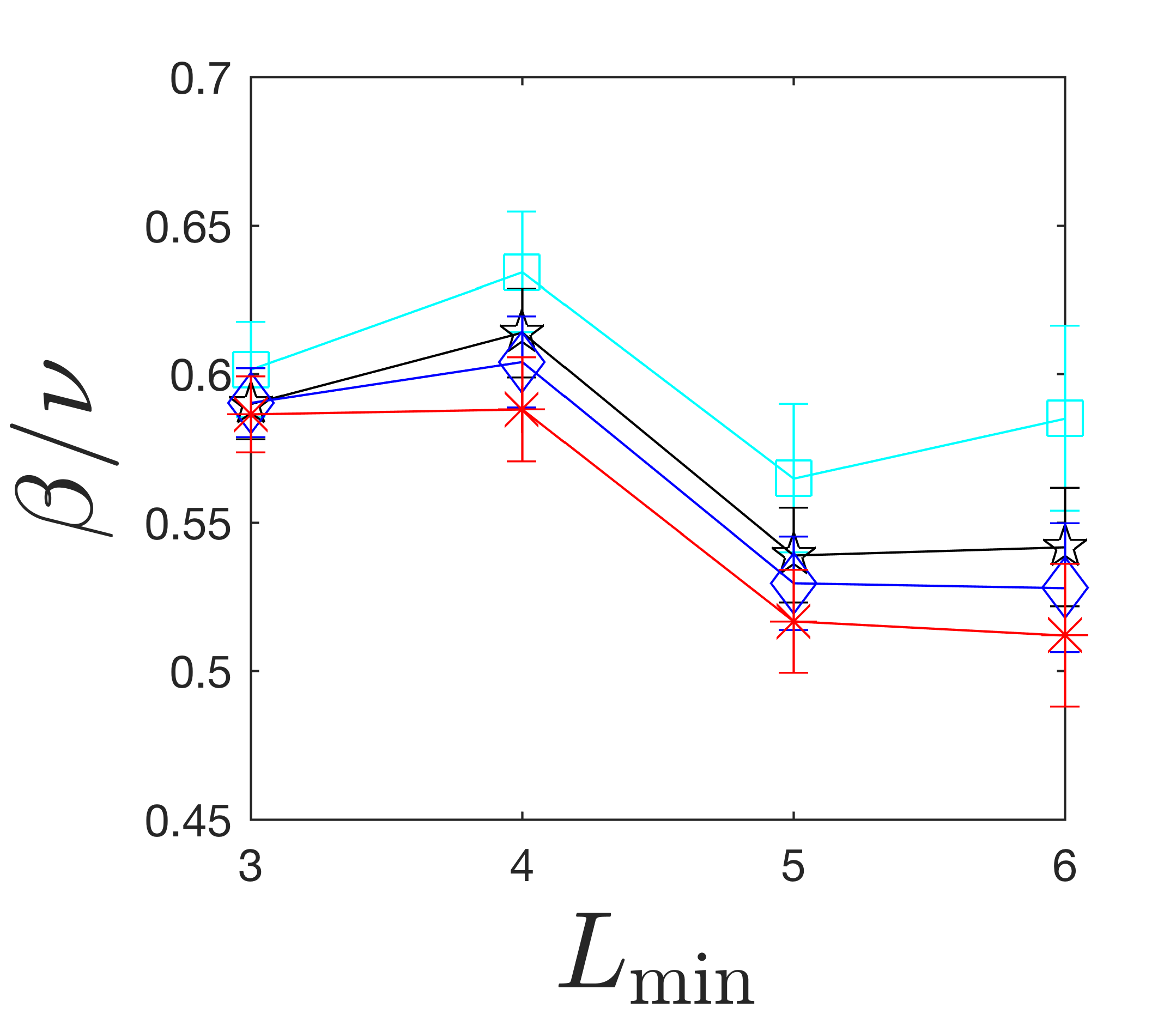}
\includegraphics[width=0.48\columnwidth]{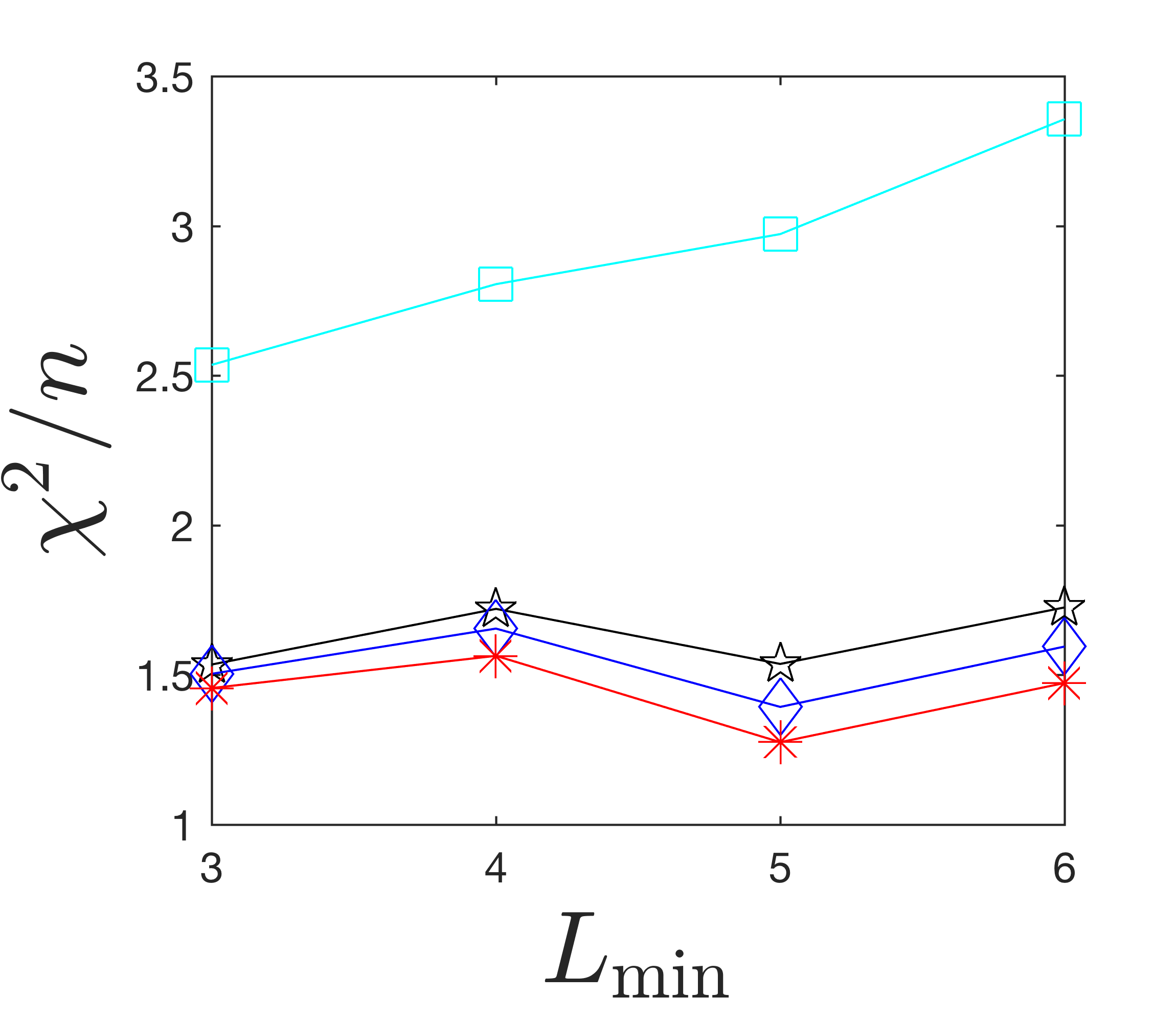}
\\ \raggedright
\caption{The critical values (a) $\nu$, (b) $\Sigma_c$, (c) $\beta/\nu$ from the Levenberg-Marquardt method for 3D simple shear are plotted versus the minimum system length for various intervals about $\Sigma_c$. Error bars represent one standard deviation. The $\chi^2 / n$ values for each fit are plotted in panel (d).}
\label{fig:crit-vals-3D}
\end{figure}

Figure~\ref{fig:crit-vals-3D} shows the critical values that yield the best fits for 3D simple shear as a function of $L_{\rm min}$ for several $X_{\rm max}$. For $X_{\rm max} = 0.3$, we find $\chi^2/n > 2.5$, signifying a poor fit. For $0.1 \leq X_{\rm max} \leq 0.2$, we find $\chi^2/n \approx 1.5$, nearly independent of $L_{\rm min}$. We estimate $\nu = 1.7 \pm 0.5$, $\Sigma_c = 0.109 \pm 0.005$ 
and $\beta / \nu = 0.57 \pm 0.07$ by the scatter in results for $0.1 \leq X_{\rm max} \leq 0.2$, plus the typical width of the error bars, which represent one standard deviation in the Levenberg-Marquardt fit. We note significant uncertainty in the value of $\nu$, which agrees with our observation that good scaling collapses are possible with $\nu$ from 1.2 to 2.2 for the 3D simple shear data.

\begin{figure}[t!!]
\raggedright
(a) \hspace{37mm} (b)  \\ 
\includegraphics[width=0.48\columnwidth]{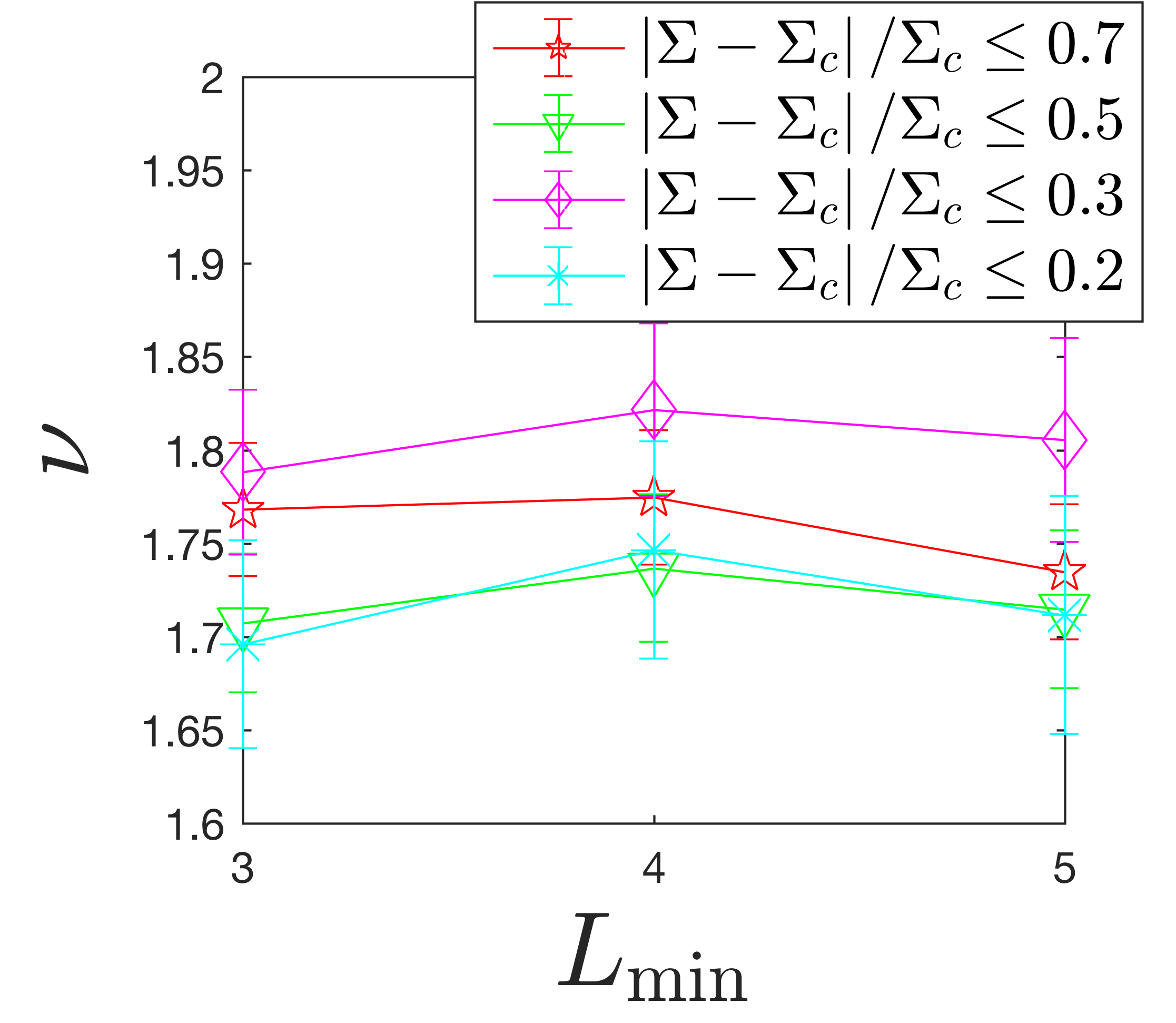}
\includegraphics[width=0.48\columnwidth]{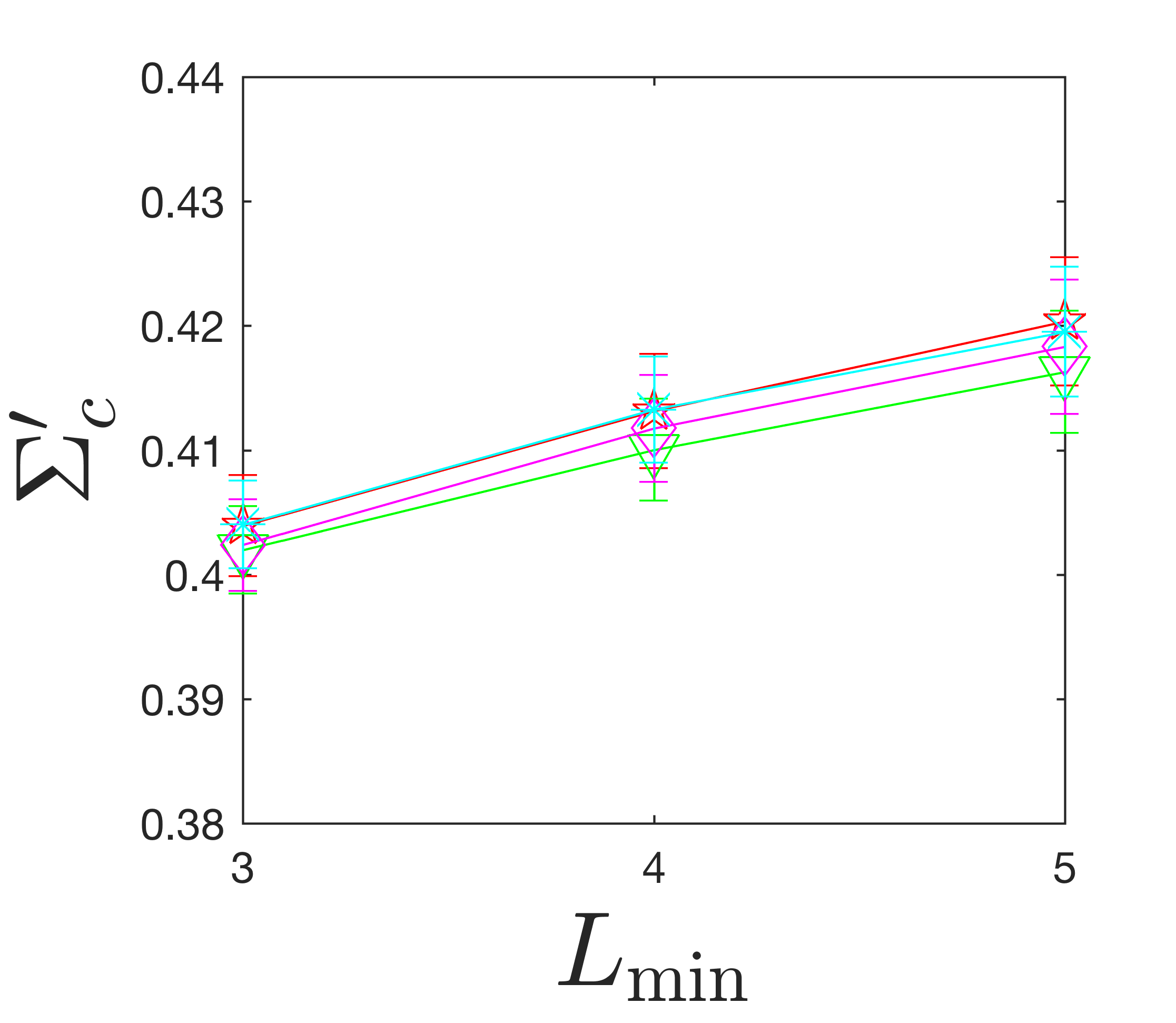}
\\ (c) \hspace{37mm} (d)  \\ 
\includegraphics[width=0.48\columnwidth]{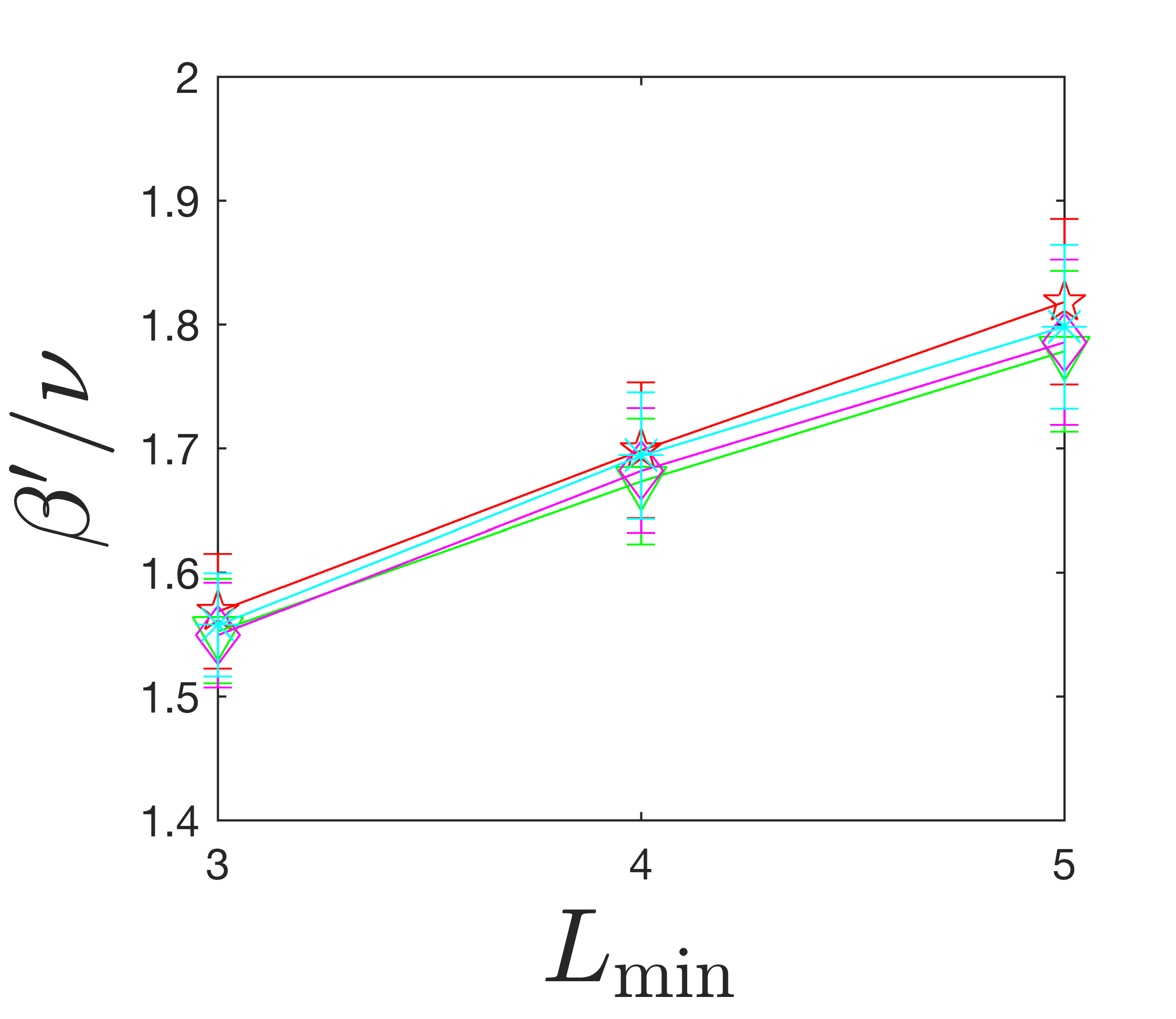}
\includegraphics[width=0.48\columnwidth]{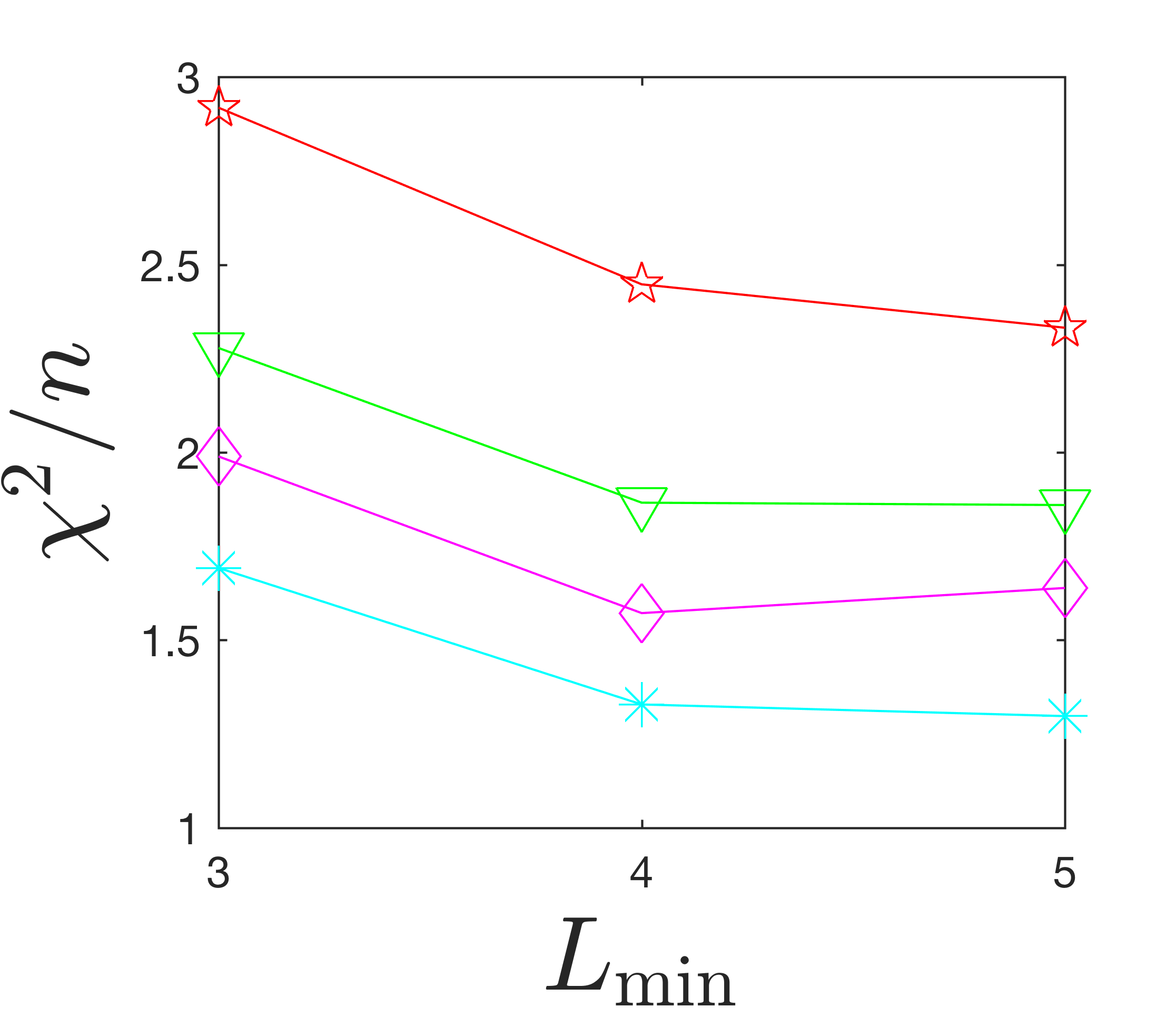}
\\ \raggedright
\caption{The critical values (a) $\nu$, (b) $\Sigma_c$, (c) $\beta/\nu$ from the Levenberg-Marquardt method for the 2D riverbed-like geometry are plotted versus the minimum system length for various intervals about $\Sigma_c$. Error bars represent one standard deviation. The $\chi^2 / n$ values for each fit are plotted in panel (d).}
\label{fig:crit-vals-RB}
\end{figure}

Figure~\ref{fig:crit-vals-RB} shows the critical values that yield the best fits for the riverbed-like geometry as a function of $L_{\rm min}$ for several values of $X_{\rm max}$. As we reduce $X_{\rm max}$, $\chi^2/n$ steadily decreases to $\approx 1$. However, the critical values are almost independent of $X_{\rm max}$ We estimate $\Sigma_c'= 0.41\pm 0.015$, $\beta'/\nu = 1.7 \pm 0.2$, and $\nu= 1.75 \pm 0.1$. A similar analysis with 2D simple shear (not shown) yields $\nu = 1.84 \pm 0.3$, $\Sigma_c = 0.11 \pm 0.01$, and $\beta/\nu \approx 0.57 \pm 0.06$. The method we describe for obtaining the critical values gives similar results to a brute force search through the parameter space, where we seek a global minimum in $\chi^2/n$. 

\section{Scaling collapse of $\langle \gamma_{\rm ms} \rangle^{-1} $ versus $\Sigma$}
\label{Appendix-L-xi}

In Fig.~\ref{fig:gamma-ms-3D}(d), we showed that the data for $\langle \gamma_{\rm ms}(\Sigma,L)\rangle^{-1}$ collapsed onto two branches when plotted as a function of the scaled system size $\tilde{L}^{-1}\equiv L^{-1}/|\Sigma-\Sigma_c|^\nu$. The inset to Fig.~\ref{fig:gamma-ms-3D}(d) showed a plot of $\langle \gamma_{\rm ms}(\Sigma,L)\rangle^{-1}$ versus $\Sigma$ for various values of $L$. We also included a curve showing the infinite-system limit for $\langle \gamma_{\rm ms}(\Sigma,L)\rangle^{-1}$ versus $\Sigma$, which is implied by the scaling in Eq.~\eqref{eq:scaling-1}.

\begin{figure}[t!!]
\raggedright 
\includegraphics[trim=0mm 0mm 0mm 0mm,clip,width=\columnwidth]{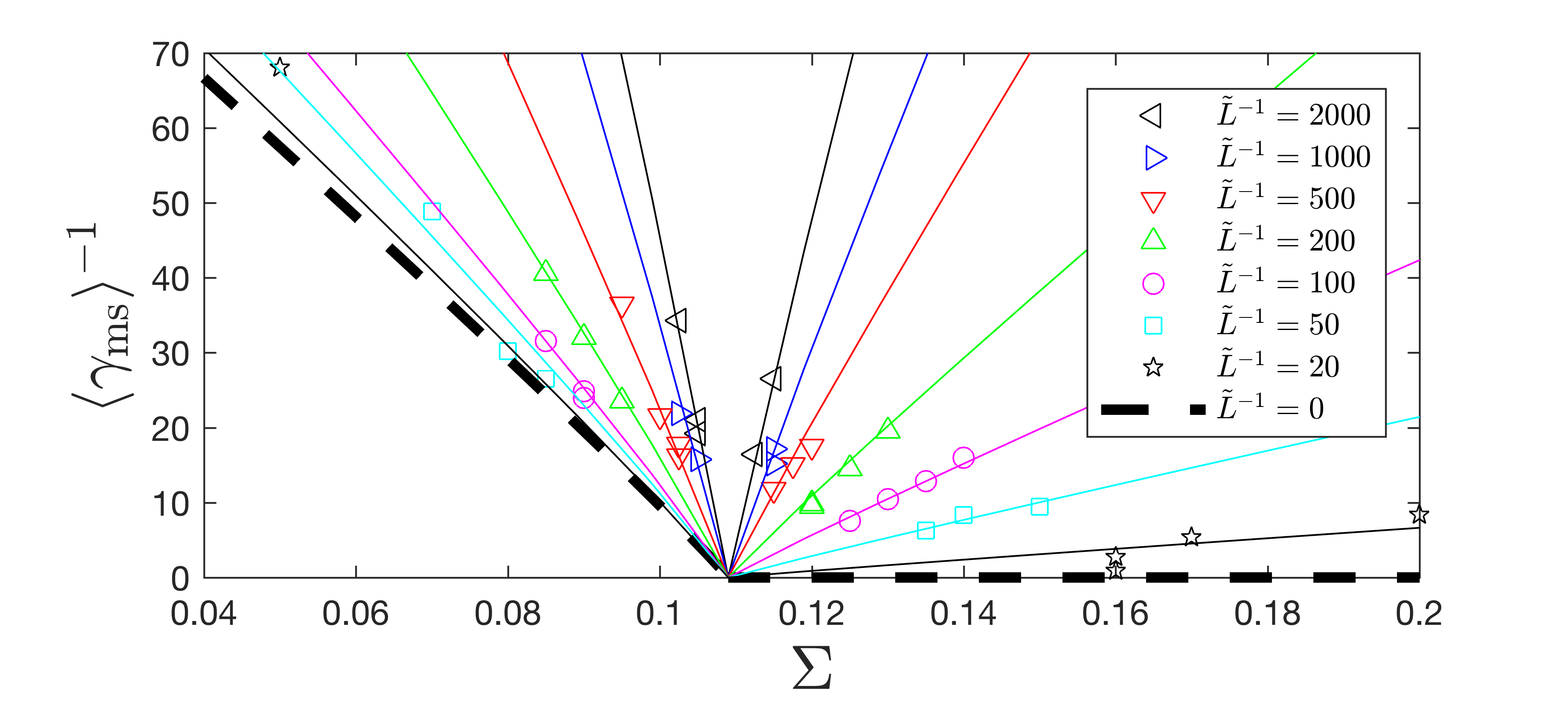}
\caption{A plot of $\langle \gamma_{\rm ms}(\Sigma,L)\rangle^{-1}$ versus $\Sigma$ for different values of $\tilde{L}^{-1}\equiv L^{-1}/|\Sigma-\Sigma_c|^\nu$. Data shown are within 10\% of the relevant value of $\tilde{L}^{-1}$ (e.g., data for $\tilde{L}^{-1}=100$ include $90<\tilde{L}^{-1}<110$). The solid curves are $\langle \gamma_{\rm ms}(\Sigma,L)\rangle^{-1} = A|\Sigma-\Sigma_c|^\beta$, where $A$ is equal to the value of $f_+$ (for $\Sigma>\Sigma_c$) or $f_-$ (for $\Sigma<\Sigma_c$) at that particular value of $\tilde{L}^{-1}$. The values of $A$ can be determined from the scaling collapse in Fig.~\ref{fig:gamma-ms-3D}(d).
}
\label{fig:order-param-L-xi}
\end{figure}

In Fig.~\ref{fig:order-param-L-xi}, we show a similar plot, but with data plotted at constant $\tilde{L}^{-1}\equiv L^{-1}/|\Sigma-\Sigma_c|^\nu$. If $\tilde{L}^{-1}$ is held fixed, Eq.~\eqref{eq:scaling-1} reduces to $\langle \gamma_{\rm ms}\rangle^{-1} = A |\Sigma - \Sigma_c|^\beta$, where $A$ is a constant, $A = f_\pm (\tilde{L}^{-1})$. The solid curves shown in Fig.~\ref{fig:order-param-L-xi} are $\langle \gamma_{\rm ms}(\Sigma,L)\rangle^{-1} = A |\Sigma - \Sigma_c|^\beta$, where the particular value of $A$ for each value of $\tilde{L}^{-1}$ at $\Sigma>\Sigma_c$ and $\Sigma<\Sigma_c$ is determined from the scaling plot in Fig.~\ref{fig:gamma-ms-3D}(d).  The solid curves pass through the data, reaffirming the scaling behavior in Eq.~\eqref{eq:scaling-1}. As $\tilde{L}^{-1}$ approaches zero, these curves approach the infinite-system limit shown by the dashed line.


\begin{acknowledgments}
This research was sponsored by the Army Research Laboratory under
Grant Numbers W911NF-14-1-0005 and W911NF-17-1-0164 (A.H.C., N.T.O.,
and C.S.O.). The views and conclusions contained in this document are
those of the authors and should not be interpreted as representing the
official policies, either expressed or implied, of the Army Research
Laboratory or the U.S. Government. The U.S. Government is authorized
to reproduce and distribute reprints for Government purposes
notwithstanding any copyright notation herein. M.D.S. also acknowledges
support from the National Science Foundation Grant No.~CMMI-1463455.
\end{acknowledgments}

\bibliography{strengthening_refs2}

\begin{thebibliography}{32}%
\makeatletter
\providecommand \@ifxundefined [1]{%
 \@ifx{#1\undefined}
}%
\providecommand \@ifnum [1]{%
 \ifnum #1\expandafter \@firstoftwo
 \else \expandafter \@secondoftwo
 \fi
}%
\providecommand \@ifx [1]{%
 \ifx #1\expandafter \@firstoftwo
 \else \expandafter \@secondoftwo
 \fi
}%
\providecommand \natexlab [1]{#1}%
\providecommand \enquote  [1]{``#1''}%
\providecommand \bibnamefont  [1]{#1}%
\providecommand \bibfnamefont [1]{#1}%
\providecommand \citenamefont [1]{#1}%
\providecommand \href@noop [0]{\@secondoftwo}%
\providecommand \href [0]{\begingroup \@sanitize@url \@href}%
\providecommand \@href[1]{\@@startlink{#1}\@@href}%
\providecommand \@@href[1]{\endgroup#1\@@endlink}%
\providecommand \@sanitize@url [0]{\catcode `\\12\catcode `\$12\catcode
  `\&12\catcode `\#12\catcode `\^12\catcode `\_12\catcode `\%12\relax}%
\providecommand \@@startlink[1]{}%
\providecommand \@@endlink[0]{}%
\providecommand \url  [0]{\begingroup\@sanitize@url \@url }%
\providecommand \@url [1]{\endgroup\@href {#1}{\urlprefix }}%
\providecommand \urlprefix  [0]{URL }%
\providecommand \Eprint [0]{\href }%
\providecommand \doibase [0]{http://dx.doi.org/}%
\providecommand \selectlanguage [0]{\@gobble}%
\providecommand \bibinfo  [0]{\@secondoftwo}%
\providecommand \bibfield  [0]{\@secondoftwo}%
\providecommand \translation [1]{[#1]}%
\providecommand \BibitemOpen [0]{}%
\providecommand \bibitemStop [0]{}%
\providecommand \bibitemNoStop [0]{.\EOS\space}%
\providecommand \EOS [0]{\spacefactor3000\relax}%
\providecommand \BibitemShut  [1]{\csname bibitem#1\endcsname}%
\let\auto@bib@innerbib\@empty
\bibitem [{\citenamefont {da~Cruz}\ \emph {et~al.}(2005)\citenamefont
  {da~Cruz}, \citenamefont {Emam}, \citenamefont {Prochnow}, \citenamefont
  {Roux},\ and\ \citenamefont {Chevoir}}]{dacruz2005}%
  \BibitemOpen
  \bibfield  {author} {\bibinfo {author} {\bibfnamefont {F.}~\bibnamefont
  {da~Cruz}}, \bibinfo {author} {\bibfnamefont {S.}~\bibnamefont {Emam}},
  \bibinfo {author} {\bibfnamefont {M.}~\bibnamefont {Prochnow}}, \bibinfo
  {author} {\bibfnamefont {J.-N.}\ \bibnamefont {Roux}}, \ and\ \bibinfo
  {author} {\bibfnamefont {F.}~\bibnamefont {Chevoir}},\ }\bibfield  {title}
  {\enquote {\bibinfo {title} {Rheophysics of dense granular materials:
  Discrete simulation of plane shear flows},}\ }\href@noop {} {\bibfield
  {journal} {\bibinfo  {journal} {Phys. Rev. E}\ }\textbf {\bibinfo {volume}
  {72}},\ \bibinfo {pages} {021309} (\bibinfo {year} {2005})}\BibitemShut
  {NoStop}%
\bibitem [{\citenamefont {Jop}\ \emph {et~al.}(2006)\citenamefont {Jop},
  \citenamefont {Forterre},\ and\ \citenamefont {Pouliquen}}]{jop2006}%
  \BibitemOpen
  \bibfield  {author} {\bibinfo {author} {\bibfnamefont {P.}~\bibnamefont
  {Jop}}, \bibinfo {author} {\bibfnamefont {Y.}~\bibnamefont {Forterre}}, \
  and\ \bibinfo {author} {\bibfnamefont {O.}~\bibnamefont {Pouliquen}},\
  }\bibfield  {title} {\enquote {\bibinfo {title} {A constitutive law for dense
  granular flows},}\ }\href@noop {} {\bibfield  {journal} {\bibinfo  {journal}
  {Nature}\ }\textbf {\bibinfo {volume} {441}},\ \bibinfo {pages} {727--730}
  (\bibinfo {year} {2006})}\BibitemShut {NoStop}%
\bibitem [{\citenamefont {Gardiner}\ \emph {et~al.}(1998)\citenamefont
  {Gardiner}, \citenamefont {Dlugogorski}, \citenamefont {Jameson},\ and\
  \citenamefont {Chhabra}}]{gardiner1998}%
  \BibitemOpen
  \bibfield  {author} {\bibinfo {author} {\bibfnamefont {B.~S.}\ \bibnamefont
  {Gardiner}}, \bibinfo {author} {\bibfnamefont {B.~Z.}\ \bibnamefont
  {Dlugogorski}}, \bibinfo {author} {\bibfnamefont {G.~J.}\ \bibnamefont
  {Jameson}}, \ and\ \bibinfo {author} {\bibfnamefont {R.~P.}\ \bibnamefont
  {Chhabra}},\ }\bibfield  {title} {\enquote {\bibinfo {title} {Yield stress
  measurements of aqueous foams in the dry limit},}\ }\href@noop {} {\bibfield
  {journal} {\bibinfo  {journal} {J. Rheol.}\ }\textbf {\bibinfo {volume}
  {42}},\ \bibinfo {pages} {1437--1450} (\bibinfo {year} {1998})}\BibitemShut
  {NoStop}%
\bibitem [{\citenamefont {Yoshimura}\ \emph {et~al.}(1987)\citenamefont
  {Yoshimura}, \citenamefont {Prud'homme}, \citenamefont {Princen},\ and\
  \citenamefont {Kiss}}]{yoshimura1987}%
  \BibitemOpen
  \bibfield  {author} {\bibinfo {author} {\bibfnamefont {A.~S}\ \bibnamefont
  {Yoshimura}}, \bibinfo {author} {\bibfnamefont {R.~K.}\ \bibnamefont
  {Prud'homme}}, \bibinfo {author} {\bibfnamefont {H.~M.}\ \bibnamefont
  {Princen}}, \ and\ \bibinfo {author} {\bibfnamefont {A.~D.}\ \bibnamefont
  {Kiss}},\ }\bibfield  {title} {\enquote {\bibinfo {title} {A comparison of
  techniques for measuring yield stresses},}\ }\href@noop {} {\bibfield
  {journal} {\bibinfo  {journal} {J. Rheol.}\ }\textbf {\bibinfo {volume}
  {31}},\ \bibinfo {pages} {699--710} (\bibinfo {year} {1987})}\BibitemShut
  {NoStop}%
\bibitem [{\citenamefont {Coussot}\ \emph {et~al.}(2002)\citenamefont
  {Coussot}, \citenamefont {Nguyen}, \citenamefont {Huynh},\ and\ \citenamefont
  {Bonn}}]{coussot2002}%
  \BibitemOpen
  \bibfield  {author} {\bibinfo {author} {\bibfnamefont {P.}~\bibnamefont
  {Coussot}}, \bibinfo {author} {\bibfnamefont {Q.~D.}\ \bibnamefont {Nguyen}},
  \bibinfo {author} {\bibfnamefont {H.~T.}\ \bibnamefont {Huynh}}, \ and\
  \bibinfo {author} {\bibfnamefont {D.}~\bibnamefont {Bonn}},\ }\bibfield
  {title} {\enquote {\bibinfo {title} {Avalanche behavior in yield stress
  fluids},}\ }\href@noop {} {\bibfield  {journal} {\bibinfo  {journal} {Phys.
  Rev. Lett.}\ }\textbf {\bibinfo {volume} {88}},\ \bibinfo {pages} {175501}
  (\bibinfo {year} {2002})}\BibitemShut {NoStop}%
\bibitem [{\citenamefont {Boromand}\ \emph {et~al.}(2017)\citenamefont
  {Boromand}, \citenamefont {Jamali},\ and\ \citenamefont
  {Maia}}]{boromand2017}%
  \BibitemOpen
  \bibfield  {author} {\bibinfo {author} {\bibfnamefont {A.}~\bibnamefont
  {Boromand}}, \bibinfo {author} {\bibfnamefont {S.}~\bibnamefont {Jamali}}, \
  and\ \bibinfo {author} {\bibfnamefont {J.~M.}\ \bibnamefont {Maia}},\
  }\bibfield  {title} {\enquote {\bibinfo {title} {Structural fingerprints of
  yielding mechanisms in attractive colloidal gels},}\ }\href@noop {}
  {\bibfield  {journal} {\bibinfo  {journal} {Soft Matter}\ }\textbf {\bibinfo
  {volume} {13}},\ \bibinfo {pages} {458--473} (\bibinfo {year}
  {2017})}\BibitemShut {NoStop}%
\bibitem [{\citenamefont {Toiya}\ \emph {et~al.}(2004)\citenamefont {Toiya},
  \citenamefont {Stambaugh},\ and\ \citenamefont {Losert}}]{toiya2004}%
  \BibitemOpen
  \bibfield  {author} {\bibinfo {author} {\bibfnamefont {Masahiro}\
  \bibnamefont {Toiya}}, \bibinfo {author} {\bibfnamefont {Justin}\
  \bibnamefont {Stambaugh}}, \ and\ \bibinfo {author} {\bibfnamefont
  {Wolfgang}\ \bibnamefont {Losert}},\ }\bibfield  {title} {\enquote {\bibinfo
  {title} {Transient and oscillatory granular shear flow},}\ }\href {\doibase
  10.1103/PhysRevLett.93.088001} {\bibfield  {journal} {\bibinfo  {journal}
  {Phys. Rev. Lett.}\ }\textbf {\bibinfo {volume} {93}},\ \bibinfo {pages}
  {088001} (\bibinfo {year} {2004})}\BibitemShut {NoStop}%
\bibitem [{\citenamefont {Xu}\ and\ \citenamefont {O'Hern}(2006)}]{xu2006}%
  \BibitemOpen
  \bibfield  {author} {\bibinfo {author} {\bibfnamefont {N.}~\bibnamefont
  {Xu}}\ and\ \bibinfo {author} {\bibfnamefont {C.~S.}\ \bibnamefont
  {O'Hern}},\ }\bibfield  {title} {\enquote {\bibinfo {title} {Measurements of
  the yield stress in frictionless granular systems},}\ }\href {\doibase
  10.1103/PhysRevE.73.061303} {\bibfield  {journal} {\bibinfo  {journal} {Phys.
  Rev. E}\ }\textbf {\bibinfo {volume} {73}},\ \bibinfo {pages} {061303}
  (\bibinfo {year} {2006})}\BibitemShut {NoStop}%
\bibitem [{\citenamefont {Peyneau}\ and\ \citenamefont
  {Roux}(2008)}]{peyneau08}%
  \BibitemOpen
  \bibfield  {author} {\bibinfo {author} {\bibfnamefont {P.-E.}\ \bibnamefont
  {Peyneau}}\ and\ \bibinfo {author} {\bibfnamefont {J.-N.}\ \bibnamefont
  {Roux}},\ }\bibfield  {title} {\enquote {\bibinfo {title} {Frictionless bead
  packs have macroscopic friction, but no dilatancy},}\ }\href {\doibase
  10.1103/PhysRevE.78.011307} {\bibfield  {journal} {\bibinfo  {journal} {Phys.
  Rev. E}\ }\textbf {\bibinfo {volume} {78}},\ \bibinfo {pages} {011307}
  (\bibinfo {year} {2008})}\BibitemShut {NoStop}%
\bibitem [{\citenamefont {O’Hern}\ \emph {et~al.}(2003)\citenamefont
  {O’Hern}, \citenamefont {Silbert}, \citenamefont {Liu},\ and\ \citenamefont
  {Nagel}}]{ohern2003}%
  \BibitemOpen
  \bibfield  {author} {\bibinfo {author} {\bibfnamefont {C.~S.}\ \bibnamefont
  {O’Hern}}, \bibinfo {author} {\bibfnamefont {L.~E.}\ \bibnamefont
  {Silbert}}, \bibinfo {author} {\bibfnamefont {A.~J.}\ \bibnamefont {Liu}}, \
  and\ \bibinfo {author} {\bibfnamefont {S.~R.}\ \bibnamefont {Nagel}},\
  }\bibfield  {title} {\enquote {\bibinfo {title} {Jamming at zero temperature
  and zero applied stress: The epitome of disorder},}\ }\href@noop {}
  {\bibfield  {journal} {\bibinfo  {journal} {Phys. Rev. E}\ }\textbf {\bibinfo
  {volume} {68}},\ \bibinfo {pages} {011306} (\bibinfo {year}
  {2003})}\BibitemShut {NoStop}%
\bibitem [{\citenamefont {Donev}\ \emph {et~al.}(2004)\citenamefont {Donev},
  \citenamefont {Torquato}, \citenamefont {Stillinger},\ and\ \citenamefont
  {Connelly}}]{donev2004}%
  \BibitemOpen
  \bibfield  {author} {\bibinfo {author} {\bibfnamefont {A.}~\bibnamefont
  {Donev}}, \bibinfo {author} {\bibfnamefont {S.}~\bibnamefont {Torquato}},
  \bibinfo {author} {\bibfnamefont {F.~H.}\ \bibnamefont {Stillinger}}, \ and\
  \bibinfo {author} {\bibfnamefont {R.}~\bibnamefont {Connelly}},\ }\bibfield
  {title} {\enquote {\bibinfo {title} {Jamming in hard sphere and disk
  packings},}\ }\href@noop {} {\bibfield  {journal} {\bibinfo  {journal} {J.
  Appl. Phys.}\ }\textbf {\bibinfo {volume} {95}},\ \bibinfo {pages} {989--999}
  (\bibinfo {year} {2004})}\BibitemShut {NoStop}%
\bibitem [{\citenamefont {Olsson}\ and\ \citenamefont
  {Teitel}(2007)}]{olsson2007}%
  \BibitemOpen
  \bibfield  {author} {\bibinfo {author} {\bibfnamefont {P.}~\bibnamefont
  {Olsson}}\ and\ \bibinfo {author} {\bibfnamefont {S.}~\bibnamefont
  {Teitel}},\ }\bibfield  {title} {\enquote {\bibinfo {title} {Critical scaling
  of shear viscosity at the jamming transition},}\ }\href@noop {} {\bibfield
  {journal} {\bibinfo  {journal} {Phys. Rev. Lett.}\ }\textbf {\bibinfo
  {volume} {99}},\ \bibinfo {pages} {178001} (\bibinfo {year}
  {2007})}\BibitemShut {NoStop}%
\bibitem [{\citenamefont {Van~Hecke}(2009)}]{vanhecke2009}%
  \BibitemOpen
  \bibfield  {author} {\bibinfo {author} {\bibfnamefont {M.}~\bibnamefont
  {Van~Hecke}},\ }\bibfield  {title} {\enquote {\bibinfo {title} {Jamming of
  soft particles: geometry, mechanics, scaling and isostaticity},}\ }\href@noop
  {} {\bibfield  {journal} {\bibinfo  {journal} {J. Phys. Condens. Matter.}\
  }\textbf {\bibinfo {volume} {22}},\ \bibinfo {pages} {033101} (\bibinfo
  {year} {2009})}\BibitemShut {NoStop}%
\bibitem [{\citenamefont {Tighe}\ \emph {et~al.}(2010)\citenamefont {Tighe},
  \citenamefont {Woldhuis}, \citenamefont {Remmers}, \citenamefont {van
  Saarloos},\ and\ \citenamefont {van Hecke}}]{Tighe2010}%
  \BibitemOpen
  \bibfield  {author} {\bibinfo {author} {\bibfnamefont {Brian~P.}\
  \bibnamefont {Tighe}}, \bibinfo {author} {\bibfnamefont {Erik}\ \bibnamefont
  {Woldhuis}}, \bibinfo {author} {\bibfnamefont {Joris J.~C.}\ \bibnamefont
  {Remmers}}, \bibinfo {author} {\bibfnamefont {Wim}\ \bibnamefont {van
  Saarloos}}, \ and\ \bibinfo {author} {\bibfnamefont {Martin}\ \bibnamefont
  {van Hecke}},\ }\bibfield  {title} {\enquote {\bibinfo {title} {Model for the
  scaling of stresses and fluctuations in flows near jamming},}\ }\href
  {\doibase 10.1103/PhysRevLett.105.088303} {\bibfield  {journal} {\bibinfo
  {journal} {Phys. Rev. Lett.}\ }\textbf {\bibinfo {volume} {105}},\ \bibinfo
  {pages} {088303} (\bibinfo {year} {2010})}\BibitemShut {NoStop}%
\bibitem [{\citenamefont {Nordstrom}\ \emph {et~al.}(2010)\citenamefont
  {Nordstrom}, \citenamefont {Verneuil}, \citenamefont {Arratia}, \citenamefont
  {Basu}, \citenamefont {Zhang}, \citenamefont {Yodh}, \citenamefont {Gollub},\
  and\ \citenamefont {Durian}}]{Nordstrom2010}%
  \BibitemOpen
  \bibfield  {author} {\bibinfo {author} {\bibfnamefont {K.~N.}\ \bibnamefont
  {Nordstrom}}, \bibinfo {author} {\bibfnamefont {E.}~\bibnamefont {Verneuil}},
  \bibinfo {author} {\bibfnamefont {P.~E.}\ \bibnamefont {Arratia}}, \bibinfo
  {author} {\bibfnamefont {A.}~\bibnamefont {Basu}}, \bibinfo {author}
  {\bibfnamefont {Z.}~\bibnamefont {Zhang}}, \bibinfo {author} {\bibfnamefont
  {A.~G.}\ \bibnamefont {Yodh}}, \bibinfo {author} {\bibfnamefont {J.~P.}\
  \bibnamefont {Gollub}}, \ and\ \bibinfo {author} {\bibfnamefont {D.~J.}\
  \bibnamefont {Durian}},\ }\bibfield  {title} {\enquote {\bibinfo {title}
  {Microfluidic rheology of soft colloids above and below jamming},}\ }\href
  {\doibase 10.1103/PhysRevLett.105.175701} {\bibfield  {journal} {\bibinfo
  {journal} {Phys. Rev. Lett.}\ }\textbf {\bibinfo {volume} {105}},\ \bibinfo
  {pages} {175701} (\bibinfo {year} {2010})}\BibitemShut {NoStop}%
\bibitem [{\citenamefont {Olsson}\ and\ \citenamefont
  {Teitel}(2011)}]{olsson2011}%
  \BibitemOpen
  \bibfield  {author} {\bibinfo {author} {\bibfnamefont {P.}~\bibnamefont
  {Olsson}}\ and\ \bibinfo {author} {\bibfnamefont {S.}~\bibnamefont
  {Teitel}},\ }\bibfield  {title} {\enquote {\bibinfo {title} {Critical scaling
  of shearing rheology at the jamming transition of soft-core frictionless
  disks},}\ }\href {\doibase 10.1103/PhysRevE.83.030302} {\bibfield  {journal}
  {\bibinfo  {journal} {Phys. Rev. E}\ }\textbf {\bibinfo {volume} {83}},\
  \bibinfo {pages} {030302} (\bibinfo {year} {2011})}\BibitemShut {NoStop}%
\bibitem [{\citenamefont {Kamrin}\ and\ \citenamefont
  {Koval}(2014)}]{kamrin2014}%
  \BibitemOpen
  \bibfield  {author} {\bibinfo {author} {\bibfnamefont {K.}~\bibnamefont
  {Kamrin}}\ and\ \bibinfo {author} {\bibfnamefont {G.}~\bibnamefont {Koval}},\
  }\bibfield  {title} {\enquote {\bibinfo {title} {Effect of particle surface
  friction on nonlocal constitutive behavior of flowing granular media},}\
  }\href@noop {} {\bibfield  {journal} {\bibinfo  {journal} {Comp. Part.
  Mech.}\ }\textbf {\bibinfo {volume} {1}},\ \bibinfo {pages} {169--176}
  (\bibinfo {year} {2014})}\BibitemShut {NoStop}%
\bibitem [{\citenamefont {Clark}\ \emph {et~al.}(2015)\citenamefont {Clark},
  \citenamefont {Shattuck}, \citenamefont {Ouellette},\ and\ \citenamefont
  {O'Hern}}]{clark2015hydro}%
  \BibitemOpen
  \bibfield  {author} {\bibinfo {author} {\bibfnamefont {A.~H.}\ \bibnamefont
  {Clark}}, \bibinfo {author} {\bibfnamefont {M.~D.}\ \bibnamefont {Shattuck}},
  \bibinfo {author} {\bibfnamefont {N.~T.}\ \bibnamefont {Ouellette}}, \ and\
  \bibinfo {author} {\bibfnamefont {C.~S.}\ \bibnamefont {O'Hern}},\ }\bibfield
   {title} {\enquote {\bibinfo {title} {Onset and cessation of motion in
  hydrodynamically sheared granular beds},}\ }\href {\doibase
  10.1103/PhysRevE.92.042202} {\bibfield  {journal} {\bibinfo  {journal} {Phys.
  Rev. E}\ }\textbf {\bibinfo {volume} {92}},\ \bibinfo {pages} {042202}
  (\bibinfo {year} {2015})}\BibitemShut {NoStop}%
\bibitem [{\citenamefont {Clark}\ \emph {et~al.}(2017)\citenamefont {Clark},
  \citenamefont {Shattuck}, \citenamefont {Ouellette},\ and\ \citenamefont
  {O'Hern}}]{clark2017}%
  \BibitemOpen
  \bibfield  {author} {\bibinfo {author} {\bibfnamefont {A.~H.}\ \bibnamefont
  {Clark}}, \bibinfo {author} {\bibfnamefont {M.~D.}\ \bibnamefont {Shattuck}},
  \bibinfo {author} {\bibfnamefont {N.~T.}\ \bibnamefont {Ouellette}}, \ and\
  \bibinfo {author} {\bibfnamefont {C.~S.}\ \bibnamefont {O'Hern}},\ }\bibfield
   {title} {\enquote {\bibinfo {title} {Role of grain dynamics in determining
  the onset of sediment transport},}\ }\href@noop {} {\bibfield  {journal}
  {\bibinfo  {journal} {Phys. Rev. Fluids}\ ,\ \bibinfo {pages} {034305}}
  (\bibinfo {year} {2017})}\BibitemShut {NoStop}%
\bibitem [{\citenamefont {Kamrin}\ and\ \citenamefont
  {Koval}(2012)}]{kamrin2012}%
  \BibitemOpen
  \bibfield  {author} {\bibinfo {author} {\bibfnamefont {K.}~\bibnamefont
  {Kamrin}}\ and\ \bibinfo {author} {\bibfnamefont {G.}~\bibnamefont {Koval}},\
  }\bibfield  {title} {\enquote {\bibinfo {title} {Nonlocal constitutive
  relation for steady granular flow},}\ }\href@noop {} {\bibfield  {journal}
  {\bibinfo  {journal} {Phys. Rev. Lett.}\ }\textbf {\bibinfo {volume} {108}},\
  \bibinfo {pages} {178301} (\bibinfo {year} {2012})}\BibitemShut {NoStop}%
\bibitem [{\citenamefont {Bouzid}\ \emph {et~al.}(2013)\citenamefont {Bouzid},
  \citenamefont {Trulsson}, \citenamefont {Claudin}, \citenamefont
  {Cl{\'e}ment},\ and\ \citenamefont {Andreotti}}]{bouzid2013}%
  \BibitemOpen
  \bibfield  {author} {\bibinfo {author} {\bibfnamefont {M.}~\bibnamefont
  {Bouzid}}, \bibinfo {author} {\bibfnamefont {M.}~\bibnamefont {Trulsson}},
  \bibinfo {author} {\bibfnamefont {P.}~\bibnamefont {Claudin}}, \bibinfo
  {author} {\bibfnamefont {E.}~\bibnamefont {Cl{\'e}ment}}, \ and\ \bibinfo
  {author} {\bibfnamefont {B.}~\bibnamefont {Andreotti}},\ }\bibfield  {title}
  {\enquote {\bibinfo {title} {Nonlocal rheology of granular flows across yield
  conditions},}\ }\href@noop {} {\bibfield  {journal} {\bibinfo  {journal}
  {Phys. Rev. Lett.}\ }\textbf {\bibinfo {volume} {111}},\ \bibinfo {pages}
  {238301} (\bibinfo {year} {2013})}\BibitemShut {NoStop}%
\bibitem [{\citenamefont {Henann}\ and\ \citenamefont
  {Kamrin}(2014)}]{henann2014}%
  \BibitemOpen
  \bibfield  {author} {\bibinfo {author} {\bibfnamefont {D.~L.}\ \bibnamefont
  {Henann}}\ and\ \bibinfo {author} {\bibfnamefont {K.}~\bibnamefont
  {Kamrin}},\ }\bibfield  {title} {\enquote {\bibinfo {title} {Continuum
  modeling of secondary rheology in dense granular materials},}\ }\href@noop {}
  {\bibfield  {journal} {\bibinfo  {journal} {Phys. Rev. Lett.}\ }\textbf
  {\bibinfo {volume} {113}},\ \bibinfo {pages} {178001} (\bibinfo {year}
  {2014})}\BibitemShut {NoStop}%
\bibitem [{\citenamefont {Kamrin}\ and\ \citenamefont
  {Henann}(2015)}]{kamrin2015}%
  \BibitemOpen
  \bibfield  {author} {\bibinfo {author} {\bibfnamefont {K.}~\bibnamefont
  {Kamrin}}\ and\ \bibinfo {author} {\bibfnamefont {D.~L.}\ \bibnamefont
  {Henann}},\ }\bibfield  {title} {\enquote {\bibinfo {title} {Nonlocal
  modeling of granular flows down inclines},}\ }\href@noop {} {\bibfield
  {journal} {\bibinfo  {journal} {Soft Matter}\ }\textbf {\bibinfo {volume}
  {11}},\ \bibinfo {pages} {179--185} (\bibinfo {year} {2015})}\BibitemShut
  {NoStop}%
\bibitem [{\citenamefont {Bouzid}\ \emph {et~al.}(2015)\citenamefont {Bouzid},
  \citenamefont {Izzet}, \citenamefont {Trulsson}, \citenamefont {Cl{\'e}ment},
  \citenamefont {Claudin},\ and\ \citenamefont {Andreotti}}]{bouzid2015}%
  \BibitemOpen
  \bibfield  {author} {\bibinfo {author} {\bibfnamefont {M.}~\bibnamefont
  {Bouzid}}, \bibinfo {author} {\bibfnamefont {A.}~\bibnamefont {Izzet}},
  \bibinfo {author} {\bibfnamefont {M.}~\bibnamefont {Trulsson}}, \bibinfo
  {author} {\bibfnamefont {E.}~\bibnamefont {Cl{\'e}ment}}, \bibinfo {author}
  {\bibfnamefont {P.}~\bibnamefont {Claudin}}, \ and\ \bibinfo {author}
  {\bibfnamefont {B.}~\bibnamefont {Andreotti}},\ }\bibfield  {title} {\enquote
  {\bibinfo {title} {Non-local rheology in dense granular flows},}\ }\href@noop
  {} {\bibfield  {journal} {\bibinfo  {journal} {Eur. Phys. J. E}\ }\textbf
  {\bibinfo {volume} {38}},\ \bibinfo {pages} {125} (\bibinfo {year}
  {2015})}\BibitemShut {NoStop}%
\bibitem [{\citenamefont {Bocquet}\ \emph {et~al.}(2009)\citenamefont
  {Bocquet}, \citenamefont {Colin},\ and\ \citenamefont
  {Ajdari}}]{bocquet2009}%
  \BibitemOpen
  \bibfield  {author} {\bibinfo {author} {\bibfnamefont {L.}~\bibnamefont
  {Bocquet}}, \bibinfo {author} {\bibfnamefont {A.}~\bibnamefont {Colin}}, \
  and\ \bibinfo {author} {\bibfnamefont {A.}~\bibnamefont {Ajdari}},\
  }\bibfield  {title} {\enquote {\bibinfo {title} {Kinetic theory of plastic
  flow in soft glassy materials},}\ }\href@noop {} {\bibfield  {journal}
  {\bibinfo  {journal} {Phys. Rev. Lett.}\ }\textbf {\bibinfo {volume} {103}},\
  \bibinfo {pages} {036001} (\bibinfo {year} {2009})}\BibitemShut {NoStop}%
\bibitem [{\citenamefont {Karmakar}\ \emph {et~al.}(2010)\citenamefont
  {Karmakar}, \citenamefont {Lerner},\ and\ \citenamefont
  {Procaccia}}]{karmakar2010}%
  \BibitemOpen
  \bibfield  {author} {\bibinfo {author} {\bibfnamefont {S.}~\bibnamefont
  {Karmakar}}, \bibinfo {author} {\bibfnamefont {E.}~\bibnamefont {Lerner}}, \
  and\ \bibinfo {author} {\bibfnamefont {I.}~\bibnamefont {Procaccia}},\
  }\bibfield  {title} {\enquote {\bibinfo {title} {Statistical physics of the
  yielding transition in amorphous solids},}\ }\href {\doibase
  10.1103/PhysRevE.82.055103} {\bibfield  {journal} {\bibinfo  {journal} {Phys.
  Rev. E}\ }\textbf {\bibinfo {volume} {82}},\ \bibinfo {pages} {055103}
  (\bibinfo {year} {2010})}\BibitemShut {NoStop}%
\bibitem [{\citenamefont {Lin}\ \emph {et~al.}(2014)\citenamefont {Lin},
  \citenamefont {Lerner}, \citenamefont {Rosso},\ and\ \citenamefont
  {Wyart}}]{lin2014}%
  \BibitemOpen
  \bibfield  {author} {\bibinfo {author} {\bibfnamefont {J.}~\bibnamefont
  {Lin}}, \bibinfo {author} {\bibfnamefont {E.}~\bibnamefont {Lerner}},
  \bibinfo {author} {\bibfnamefont {A.}~\bibnamefont {Rosso}}, \ and\ \bibinfo
  {author} {\bibfnamefont {M.}~\bibnamefont {Wyart}},\ }\bibfield  {title}
  {\enquote {\bibinfo {title} {Scaling description of the yielding transition
  in soft amorphous solids at zero temperature},}\ }\href@noop {} {\bibfield
  {journal} {\bibinfo  {journal} {Proc. Natl. Acad. Sci.}\ }\textbf {\bibinfo
  {volume} {111}},\ \bibinfo {pages} {14382--14387} (\bibinfo {year}
  {2014})}\BibitemShut {NoStop}%
\bibitem [{\citenamefont {Xu}\ \emph {et~al.}(2005)\citenamefont {Xu},
  \citenamefont {O'Hern},\ and\ \citenamefont {Kondic}}]{xu2005}%
  \BibitemOpen
  \bibfield  {author} {\bibinfo {author} {\bibfnamefont {N.}~\bibnamefont
  {Xu}}, \bibinfo {author} {\bibfnamefont {C.~S.}\ \bibnamefont {O'Hern}}, \
  and\ \bibinfo {author} {\bibfnamefont {L.}~\bibnamefont {Kondic}},\
  }\bibfield  {title} {\enquote {\bibinfo {title} {Velocity profiles in
  repulsive athermal systems under shear},}\ }\href@noop {} {\bibfield
  {journal} {\bibinfo  {journal} {Phys. Rev. Lett.}\ }\textbf {\bibinfo
  {volume} {94}},\ \bibinfo {pages} {016001} (\bibinfo {year}
  {2005})}\BibitemShut {NoStop}%
\bibitem [{\citenamefont {Bertrand}\ \emph {et~al.}(2016)\citenamefont
  {Bertrand}, \citenamefont {Behringer}, \citenamefont {Chakraborty},
  \citenamefont {O'Hern},\ and\ \citenamefont {Shattuck}}]{bertrand2016}%
  \BibitemOpen
  \bibfield  {author} {\bibinfo {author} {\bibfnamefont {T.}~\bibnamefont
  {Bertrand}}, \bibinfo {author} {\bibfnamefont {R.~P.}\ \bibnamefont
  {Behringer}}, \bibinfo {author} {\bibfnamefont {B.}~\bibnamefont
  {Chakraborty}}, \bibinfo {author} {\bibfnamefont {C.~S.}\ \bibnamefont
  {O'Hern}}, \ and\ \bibinfo {author} {\bibfnamefont {M.~D.}\ \bibnamefont
  {Shattuck}},\ }\bibfield  {title} {\enquote {\bibinfo {title} {Protocol
  dependence of the jamming transition},}\ }\href {\doibase
  10.1103/PhysRevE.93.012901} {\bibfield  {journal} {\bibinfo  {journal} {Phys.
  Rev. E}\ }\textbf {\bibinfo {volume} {93}},\ \bibinfo {pages} {012901}
  (\bibinfo {year} {2016})}\BibitemShut {NoStop}%
\bibitem [{\citenamefont {Bi}\ \emph {et~al.}(2011)\citenamefont {Bi},
  \citenamefont {Zhang}, \citenamefont {Chakraborty},\ and\ \citenamefont
  {Behringer}}]{bi2011}%
  \BibitemOpen
  \bibfield  {author} {\bibinfo {author} {\bibfnamefont {D.}~\bibnamefont
  {Bi}}, \bibinfo {author} {\bibfnamefont {J.}~\bibnamefont {Zhang}}, \bibinfo
  {author} {\bibfnamefont {B.}~\bibnamefont {Chakraborty}}, \ and\ \bibinfo
  {author} {\bibfnamefont {R.~P.}\ \bibnamefont {Behringer}},\ }\bibfield
  {title} {\enquote {\bibinfo {title} {Jamming by shear},}\ }\href@noop {}
  {\bibfield  {journal} {\bibinfo  {journal} {Nature}\ }\textbf {\bibinfo
  {volume} {480}},\ \bibinfo {pages} {355--358} (\bibinfo {year}
  {2011})}\BibitemShut {NoStop}%
\bibitem [{\citenamefont {Baity-Jesi}\ \emph {et~al.}(2017)\citenamefont
  {Baity-Jesi}, \citenamefont {Goodrich}, \citenamefont {Liu}, \citenamefont
  {Nagel},\ and\ \citenamefont {Sethna}}]{baity2017}%
  \BibitemOpen
  \bibfield  {author} {\bibinfo {author} {\bibfnamefont {M.}~\bibnamefont
  {Baity-Jesi}}, \bibinfo {author} {\bibfnamefont {C.~P.}\ \bibnamefont
  {Goodrich}}, \bibinfo {author} {\bibfnamefont {A.~J.}\ \bibnamefont {Liu}},
  \bibinfo {author} {\bibfnamefont {S.~R.}\ \bibnamefont {Nagel}}, \ and\
  \bibinfo {author} {\bibfnamefont {J.~P.}\ \bibnamefont {Sethna}},\ }\bibfield
   {title} {\enquote {\bibinfo {title} {Emergent \textit{SO}(3) symmetry of the
  frictionless shear jamming transition},}\ }\href@noop {} {\bibfield
  {journal} {\bibinfo  {journal} {J. Stat. Phys.}\ }\textbf {\bibinfo {volume}
  {167}},\ \bibinfo {pages} {735--748} (\bibinfo {year} {2017})}\BibitemShut
  {NoStop}%
\bibitem [{\citenamefont {Chen}\ \emph {et~al.}(2018)\citenamefont {Chen},
  \citenamefont {Jin}, \citenamefont {Bertrand}, \citenamefont {Shattuck},\
  and\ \citenamefont {O'Hern}}]{chen2018}%
  \BibitemOpen
  \bibfield  {author} {\bibinfo {author} {\bibfnamefont {S.}~\bibnamefont
  {Chen}}, \bibinfo {author} {\bibfnamefont {W.}~\bibnamefont {Jin}}, \bibinfo
  {author} {\bibfnamefont {T.}~\bibnamefont {Bertrand}}, \bibinfo {author}
  {\bibfnamefont {M.~D.}\ \bibnamefont {Shattuck}}, \ and\ \bibinfo {author}
  {\bibfnamefont {C.~S.}\ \bibnamefont {O'Hern}},\ }\bibfield  {title}
  {\enquote {\bibinfo {title} {Stress anisotropy in shear-jammed packings of
  frictionless disks},}\ }\href@noop {} {\bibfield  {journal} {\bibinfo
  {journal} {arXiv:1804.10962}\ } (\bibinfo {year} {2018})}\BibitemShut
  {NoStop}%
\end{thebibliography}%

\end{document}